\documentclass[twocolumn, tighten, times]{aastex701}

\usepackage{xspace,amsmath,amssymb}
\usepackage{xcolor, fontawesome}
\usepackage{tabularx}
\usepackage[utf8]{inputenc}
\usepackage{textcomp} % provide euro and other symbols
\usepackage[capitalize]{cleveref}
\usepackage{booktabs}

\newcommand{\Msun}{\mbox{$M_{\odot}$}}
\newcommand{\Lsun}{\mbox{$L_{\odot}$}}
\newcommand{\kepler}{{\em Kepler}}
\newcommand{\cd}{\mbox{d$^{-1}$}}

\newcommand{\vsini}{\ensuremath{v\sin i}}
\newcommand{\teff}{\ensuremath{T_\mathrm{eff}}}
\newcommand{\ionn}[2]{\textup{#1\,\textsc{\uppercase{#2}}}}
\newcommand{\logg}{\ensuremath{\mathrm{log}\ g}}

\newcommand{\tess}{\mbox{\textit{TESS}}\xspace}
\newcommand{\dsct}{\mbox{$\delta$~Sct}\xspace}
\newcommand{\vbroad}{\texttt{vbroad}\xspace}

\newcommand{\samplesize}{1{,}526{,}218}      % parent A/F sample after cuts
\newcommand{\ndsct}{{156{,}706}}        % classified dSct pulsators

\newcommand{\rev}[1]{{\color{black}{#1}}}

\shorttitle{TESS $\delta$ Scuti stars}
\shortauthors{Hey et. al}

\graphicspath{{./}{figures/}}

\begin{document}

\title{The $\delta$ Scuti instability strip and period--luminosity relation as seen by TESS and Gaia DR3}

\author[0000-0003-3244-5357]{Daniel Hey}
\affiliation{Institute for Astronomy, University of Hawai`i, 2680 Woodlawn Drive, Honolulu, HI 96822, USA}
\email[show]{dhey@hawaii.edu}

\author[0000-0002-5648-3107]{Simon Murphy}
\affiliation{Centre for Astrophysics, University of Southern Queensland, Toowoomba, QLD 4350, Australia}
\email{Simon.Murphy@usq.edu.au}

\author[0000-0001-5222-4661]{Timothy R. Bedding}
\affiliation{Sydney Institute for Astronomy, School of Physics, University of Sydney, NSW 2006, Australia}
\email{tim.bedding@sydney.edu.au}

\author[0000-0003-0619-0207]{Anuj Gautam}
\affiliation{Centre for Astrophysics, University of Southern Queensland, Toowoomba, QLD 4350, Australia}
\email{anuj.gautam@unisq.edu.au}

\author[0000-0001-8832-4488]{Daniel Huber}
\affiliation{Institute for Astronomy, University of Hawai`i, 2680 Woodlawn Drive, Honolulu, HI 96822, USA}
\email{huberd@hawaii.edu}

\author[0000-0003-3020-4437]{Yaguang Li}
\affiliation{Institute for Astronomy, University of Hawai`i, 2680 Woodlawn Drive, Honolulu, HI 96822, USA}
\email{yaguangl@hawaii.edu}

\author[0009-0009-3641-4137]{Ian Berry}
\affiliation{Institute for Astronomy, University of Hawai`i, 2680 Woodlawn Drive, Honolulu, HI 96822, USA}
\email{ianberry@hawaii.edu}

\author[0009-0009-6602-4703]{Hyde Kenney}
\affiliation{Department of Astronomy, University of Florida, Gainesville, FL 32611, USA}
\affiliation{Institute for Astronomy, University of Hawai`i, 2680 Woodlawn Drive, Honolulu, HI 96822, USA}
\email{hkenney@ufl.edu}

\author[0000-0002-8707-201X]{Prasad Mani}
\affiliation{Sydney Institute for Astronomy, School of Physics, University of Sydney, NSW 2006, Australia}
\email{prasadmani94@gmail.com}

\author[0009-0000-7419-8118]{Tom Love}
\affiliation{Centre for Astrophysics, University of Southern Queensland, Toowoomba, QLD 4350, Australia}
\email{tom.love@unisq.edu.au}

\author[0000-0002-3334-9984]{Mykyta Kliapets}
\affiliation{Institute of Astronomy, KU Leuven, Celestijnenlaan 200D, bus 2401, 3001 Leuven, Belgium}
\email{mykyta.kliapets@kuleuven.be}

\begin{abstract}
\noindent 
We present an asteroseismic survey of \samplesize\ A- and F-type stars in and around the \dsct\ instability strip using TESS 10-minute cadence data, which resolve the full \dsct\ pulsation spectra across nearly the entire sky. We derive homogeneous stellar parameters from \textit{Gaia} DR3 parallaxes and archival photometry, with temperatures that agree with interferometry to within $\sim$300\,K. A Positive--Unlabelled classifier identifies \ndsct\ \dsct\ pulsators, which on average rotate significantly faster than non-variable A-type stars. Within the instability strip in the $\teff$--$\log L$ plane, the pulsator fraction peaks at {$\sim 30\%$} over the full sample and rises to {$\sim 80\%$} in the brightest magnitude bins. Detection sensitivity therefore accounts for much, but not all, of the apparent impurity of the strip. Our period--luminosity relation agrees with previous determinations, and we recast it as a scale-invariant period--density relation that connects directly to the pulsation constant. We find that the secondary ridge in these relations is probably a combination of several overtone modes. The \dsct\ stars closest to the fundamental ridge have systematically lower rotational velocities and occupy a distinct region of the instability strip. Finally, a cross-match with a \textit{Gaia} catalog of open clusters reveals a correlation between the mean pulsation frequency of \dsct\ stars and cluster age, suggesting a new ensemble asteroseismic age indicator for young stellar populations.

\end{abstract}


\section{Introduction}
\label{sec:intro}

The $\delta$ Scuti variables are intermediate-mass (1.5--2.3\,\Msun) main-sequence stars located within the classical instability strip \citep{Breger1979Delta, Breger2000Scuti, Uytterhoeven2011Kepler}. They exhibit low radial-order pressure (p) modes driven primarily by the $\kappa$ mechanism believed to be acting in the second helium ionisation zone, with pulsation frequencies typically spanning 5--70\,\cd\ and effective temperatures in the range 6500--10\,000\,K \citep{Houdek2000Convective, Antoci2019First, Kurtz2022Asteroseismology}. As intermediate-mass main-sequence pulsators, the \dsct\ stars are progenitors of the bulk of the white-dwarf mass distribution \citep{Cummings2018White} and provide an important link between the well-studied solar-like oscillators and the more massive $\beta$ Cephei and slowly pulsating B stars.

Despite their potential as asteroseismic probes, the \dsct\ stars have long resisted systematic mode identification routinely applied to the solar-like oscillators. In stars cooler than the Kraft break \citep{Kraft1967Studies, Beyer2024Kraft}, stochastic excitation by near-surface convection drives (and damps) the full range of modes within a resonant envelope, producing the regular comb of frequencies used for solar-like asteroseismology \citep[e.g.][]{Chaplin2011Ensemble}. Pulsations in \dsct\ stars, by contrast, are driven by a heat-engine mechanism and only an apparently pseudo-random subset of modes are excited across much wider frequency ranges. Their mode selection mechanism thus remains poorly understood \citep{bregerModeSelectionScuti2009, Bowman2018Characterizing}.

Mode identification is further complicated by the rapid rotation typical of A-type stars \citep{Zorec2012Rotational}, which lifts the degeneracy of azimuthal mode components and produces dense, asymmetric splitting \citep{Aerts2010Asteroseismology, Aerts2024Asteroseismic, Guo2024Oscillation}. Only with the high-cadence, all-sky coverage of the Transiting Exoplanet Survey Satellite (TESS; \citealp{Ricker2015Transiting}) mission \rev{has} a subset of so-called ``nice'' \dsct\ variables been identified that show regular, identifiable modes with a clear frequency spacing ($\Delta\nu$) that is a good diagnostic of stellar density \citep{Bedding2020Very}. These targets usually lie close to the zero-age main sequence and pulsate at high frequencies ($\geq$ 40 d$^{-1}$), and they have provided the first robust mode identifications in this class, marking a pathway towards the asteroseismic exploitation of the larger population of \dsct\ stars with more complicated pulsation spectra.

A closely related question is whether all stars within the \dsct\ instability strip pulsate and, if not, what mechanisms distinguish the pulsators from their quiescent counterparts. It is now well established, largely through the {\em Kepler\/} mission, that the instability strip is not pure: even after accounting for the reduced detection sensitivity at fainter magnitudes, the pulsator fraction peaks at only {${\sim}70\%$} at the centre of the observational strip \citep{Murphy2019Gaiaderived}, and around half of the stars in the \dsct\ temperature range show no detectable pulsation \citep{balonaKeplerObservationsScuti2011, Balona2015Pulsation}. Conversely, a small population of \dsct\ variables has recently been identified lying outside the classical strip \citep{Thomson-Paressant2025Discovery}, blurring its boundaries in the opposite sense. A limitation of the large {\em Kepler\/} population studies was that most A/F targets were observed at 30-minute cadence, placing the nominal Nyquist frequency at $\approx 24$\,\cd, well below the upper end of the \dsct\ pulsation range. The slightly irregular observing cadence of {\em Kepler}, once barycentric time corrections had been made, allowed intrinsic super-Nyquist frequencies to be distinguished from their aliases \citep{Murphy2013SuperNyquist, Shibahashi2018SuperNyquist, Hey2019Six}, although the non-zero integration times attenuated their observed amplitudes and complicated population-level selection and amplitude comparisons.

Subsequent work has highlighted the dominant role played by rotation in shaping the \dsct\ population. \citet{Gootkin2024New} classified ${\sim}16{,}000$ \dsct\ variables using \tess\ 30-minute FFI photometry and showed that the pulsator fraction within the instability strip is a strong function of rotational velocity, traced by the \textit{Gaia} \vbroad\ line-broadening parameter. \citet{Murphy2024$d$} and \citet{Wang2025Rotation} reached a similar conclusion using independent data. Recently, \citet{Berry2026Occurrence} further suggest a correlated age dependence. These findings are consistent with the expectation that the partial ionisation zone responsible for the $\kappa$ mechanism is more efficiently maintained against gravitational settling in rapidly rotating stars, providing an explanation for the lower pulsator fraction among the slowly rotating, chemically peculiar Am stars \citep{Breger1979Delta, Kurtz1989highovertone, Durfeldt-Pedros2024Variability}. However, the existence of a small population of slowly-rotating \dsct\ pulsators within the instability strip remains poorly understood, pointing towards an alternative excitation mechanism such as the turbulent pressure invoked by \citet{Antoci2014Role}. 
% Most recently, \citet{Guo2024Oscillation} have shown that the rotational splittings of low-degree modes in rapidly rotating \dsct\ and $\beta$ Cephei stars are themselves asymmetric and systematically dependent on angular degree as a consequence of non-spherical distortion caused by such rotation, pointing to a new pathway towards mode identification.

\begin{figure}
    \centering
    \includegraphics[width=\linewidth]{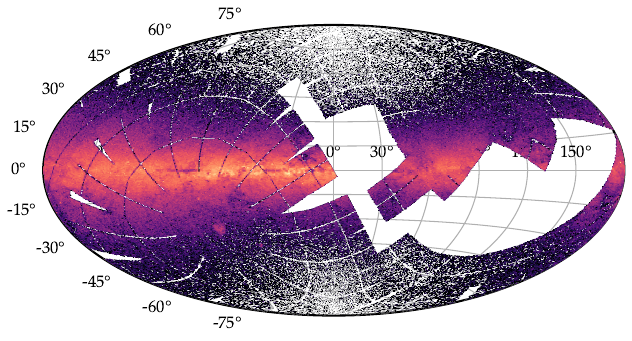}
    \caption{Sky distribution of the A/F sample (Section~\ref{ssec:targets}) in Galactic coordinates, colored by the density of points. Note the dearth of targets in certain regions caused by the absence of TESS pointings in Sectors 27 through 55 used in this study.}
    \label{fig:sky_sample}
\end{figure}

The \tess\ mission now provides the data needed to address these questions on a homogeneous basis. The 10-minute FFI cadence available from \tess\ Cycle\,3 (Sector\,27) onwards \citep{Jenkins2021Upping} places the Nyquist frequency at 72\,\cd, bringing most of the \dsct\ pulsation range below the nominal Nyquist limit and substantially reducing the aliasing encountered in 30-minute data. Combining with the precise parallaxes and photometry of \textit{Gaia} DR3 \citep{Collaboration2023Gaia} allows us to calculate luminosities and effective temperatures for over a million A and F stars across the sky to a precision previously available only for the small \textit{Kepler} sample.
% \citet{Zhou2025Identification, Zhou2025Systematic} have compiled a large vetted catalogue of \tess\ \dsct\ stars, providing a positively-labelled training set well suited to population-level classification.

In this paper, we combine \tess\ 10-minute FFI photometry from Sectors 27 through 55 (Cycle 3 and 4) with \textit{Gaia} DR3 parallaxes to construct a nearly all-sky sample of \samplesize\ A- and F-type stars (Figure~\ref{fig:sky_sample}). We use this sample to characterise the \dsct\ population across the HRD, investigate the period--luminosity relation, investigate the effects of rotation, and produce a catalogue of \dsct\ variables in open clusters. 

% In Section~\ref{sec:obs} we describe our target selection and \tess\ photometry. In Section~\ref{sec:stellar_props} we  derive stellar parameters and validate our temperature scale against interferometric measurements. In Section~\ref{sec:pulsators} we classify \dsct\ pulsators with a Positive--Unlabelled learning framework. Section~\ref{sec:results} presents the H--R diagram, the pulsator fraction, and the influence of rotation; Section~\ref{sec:pl} presents the period--luminosity and period--density relations; and Section~\ref{sec:clusters} presents the dependence of mean pulsation frequency on cluster age.

% ===============================================================
\section{Data}
\label{sec:obs}

\subsection{Target selection}
\label{ssec:targets}

We selected our targets from the \tess\ Input Catalog (TIC; \citealp{Stassun2018TESS}) because of its all-sky homogeneity. We applied a temperature cut of $6300 \le \teff \le 10{,}000$\,K and a magnitude cut of $T < 13.5$\,mag. The temperature range encompasses the classical $\delta$\,Sct instability strip \citep{Murphy2019Gaiaderived, dupretTheoreticalInstabilityStrips2004} with a buffer of $\sim$200\,K on the cool side, sufficient to capture pulsators whose catalogued temperatures are systematically biased. Stars hotter than $10{,}000$\,K are predominantly B-type pulsators ($\beta$\,Cep, SPB) and lie outside the scope of this work, while genuine \dsct\ stars cooler than $6300$\,K are rare and probably result from inaccurate photometric temperatures.

We further required each target to have a \textit{Gaia} DR3 source match with a fractional parallax uncertainty $\sigma_\pi / \pi < 0.2$, which removed 12,714 targets and ensured that distances (and hence luminosities) are well constrained. Applying these cuts yielded a parent sample of \samplesize\ A- and F-type stars, shown in Figure~\ref{fig:sky_sample}.

\subsection{\tess\ photometry}
\label{ssec:tess}

We used \tess\ 10-minute full-frame image (FFI) light curves, which became available from Cycle\,3 (Sector\,27) onwards \citep{Ricker2015Transiting, Jenkins2021Upping}. The 10-min cadence places the Nyquist frequency at $72$\,\cd, bringing most of the expected \dsct\ pulsation range below the nominal Nyquist limit and greatly reducing the aliasing affecting \kepler\ long-cadence studies \citep{Murphy2013SuperNyquist, Shibahashi2018SuperNyquist, Hey2019Six} and earlier \tess\ studies based on 30-minute FFIs \citep{Gootkin2024New}. Although shorter-cadence FFIs of 200\,s are now available from Cycle 5 onwards, we restricted our work to the 10-min FFIs in Cycles 3 and 4 in order to produce a homogeneous all-sky sample.

We used light curves produced by the \textsc{TGLC} pipeline \citep[TESS--Gaia Light Curve][]{Han2023TESSGaia}, which performs PSF photometry on the \tess\ FFIs using a forward model of the local stellar field constructed using \textit{Gaia} DR3 as a prior. For each target, we stitched together all available 10-minute FFI sectors and applied a high-pass filter to remove instrumental systematics and long-term trends. This high-pass filtering was achieved by smoothing the light curve with a Gaussian with FWHM $1$\,\cd\ and subtracting this smoothed version from the original. We utilized the internal cross-match provided in the TGLC bulk download scripts, which maps the \textit{Gaia} DR3 IDs to a corresponding TIC ID. When this work was done, bulk downloads for the TGLC photometry were   only available up to Sector 58. Because of this, we restricted our analysis to Sectors 27 through 55 inclusive, covering TESS cycles 3 and 4.

\rev{The \tess pixel scale is 21\arcsec, so light from neighbouring stars can enter the photometric aperture of a given target. This is of some concern for our sample because the majority of it lies close to the Galactic plane (Section~\ref{ssec:gaia}). The flux of a neighbour can both dilute the variability of a target and raise the noise level. Likewise, since \textsc{TGLC} light curves model neighbours as constant sources, any nearby variability can also enter the aperture of the target. We quantify these effects, and test how our results depend on Galactic latitutde in Appendix~\ref{app:blending}. From this, we identify probable blends in Table~\ref{tab:params}, and find that excluding them does not change our results.}

We computed the Lomb--Scargle periodogram \citep{Lomb1976Leastsquares, Scargle1982Studies, VanderPlas2018Understanding} of each light curve up to the nominal Nyquist frequency of $72$\,\cd. This provides unaliased frequency measurements for most \dsct\ modes, although modes intrinsically above $72$\,\cd\ remain aliased and the amplitudes of modes approaching the Nyquist limit are attenuated by the non-zero integration time.
% We adopt the $95$th-percentile Fourier amplitude as our noise estimator, following \citet{Murphy2019Gaiaderived}, and use this in the signal-to-noise calculations of Section~\ref{sec:pulsators}.

\subsection{\textit{Gaia} DR3 parallaxes and photometry}
\label{ssec:gaia}

For each target we retrieved \textit{Gaia} DR3 parallaxes, broadband $G$, $G_{\rm BP}$, and $G_{\rm RP}$ photometry, and the \vbroad\ line-broadening parameter where available \citep{Collaboration2023Gaia}. We found that 94\% of our sample have $\sigma_\pi/\pi < 0.05$. \rev{The distances of the parent sample range from 8\,pc to ${\sim}18$\,kpc, with a median of 820\,pc and 90\% of the stars lying between 290 and 2250\,pc. The \dsct\ pulsators identified in Section~\ref{sec:pulsators} follow a similar distribution, with a median distance of 860\,pc and 90\% between 250 and 1920\,pc. At these distances the A and F stars are concentrated towards the Galactic disk, with around 60\% of the parent sample lying within $10\arcdeg$ of the Galactic plane (Figure~\ref{fig:sky_sample}).} We made no further selection on photometric quality but we did require that the \textit{Gaia} DR3 \texttt{ruwe} flag was $< 1.4$, to exclude probable astrometric binaries whose parallaxes may be biased \citep{Lindegren2018Renormalising}, which excludes 16.8\% of our sample. While this high RUWE sample is excluded from analysis (Sec.~\ref{sec:results} onwards), it is still present in our final table \rev{(Table~\ref{tab:params})} and flagged accordingly.

The \vbroad\ parameter was available for 434,987 of our targets. This provides a measure of spectral line broadening that combines contributions from rotation, micro- and macroturbulence, and unresolved binarity. While not exactly equivalent to the projected rotational velocity, \vbroad\ has been shown to correlate well with \vsini\ for A- and F-type stars \citep{Fremat2023Gaia}. We note that \vbroad\ carries known temperature- and magnitude-dependent systematics and is least reliable at low broadening \citep{Fremat2023Gaia}, however \citet{Murphy2024$d$} showed that it is moderately reliable for the A/F stars in most rotation regimes.
                            
% ===============================================================
% ===============================================================
\section{Stellar properties}
\label{sec:stellar_props}

We derived homogeneous stellar properties for our full sample using the \textsc{isoclassify} package \citep{Huber2017Asteroseismology}, which has been validated against asteroseismic and interferometric benchmarks for FGK stars \citep{Berger2020GaiaKepler, Berger2026GaiaKeplerTESShost}. Here, we extend its use to the considerably hotter A- and early F-type regime, and verify the reliability of the stellar parameters using interferometry.

\textsc{isoclassify} provides two modes of operation. In \emph{direct} mode, effective temperatures are derived from empirical surface-brightness--color relations \citep{Casagrande2010Absolutely, Graczyk2017Surface}, and the stellar radius is then obtained directly from the angular diameter, parallax, and bolometric flux. In \emph{grid} mode, observed quantities (parallax, apparent magnitudes, spectroscopic constraints, etc.) are compared against a grid of stellar isochrones, and posterior distributions over \teff, \logg, mass, radius, and age are obtained by marginalising over the grid. The two modes are complementary: direct mode is largely model-independent and is the preferred choice for cool, well-calibrated stars, while grid mode incorporates stellar-evolution physics and performs better in regimes where the color--temperature relations are poorly calibrated.

For our sample of A- and F-type stars, direct mode is expected to perform poorly. The peak of the spectral energy distribution of an A-type star lies in the ultraviolet, so optical and near-infrared colors sample the Rayleigh--Jeans tail of the Planck function and become only weakly sensitive to effective temperature. The empirical surface-brightness relations on which direct mode relies are also predominantly calibrated using FGK stars, with relatively few hot calibrators available. We therefore expect direct mode to systematically degrade in accuracy with increasing temperature, and we therefore used grid mode for our primary analysis.

\subsection{Validation against interferometry}
\label{ssec:interf_validation}

\begin{figure}
\includegraphics[width=\linewidth]{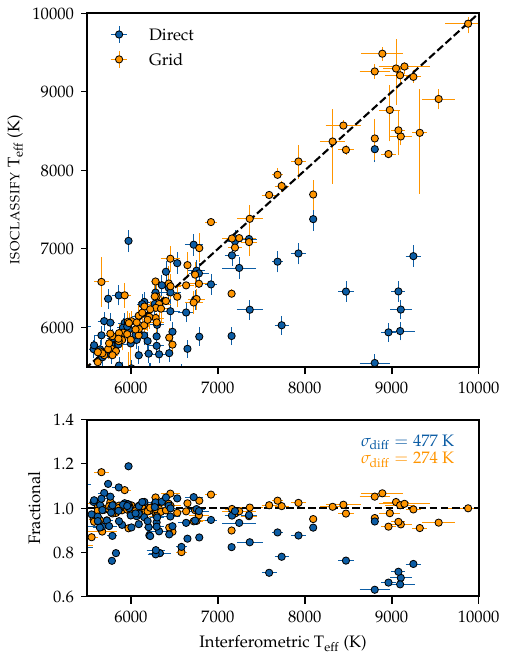}
\caption{Comparison of effective temperatures derived with the \textsc{isoclassify} direct (blue) and grid (orange) modes against interferometric temperatures from the CHARA sample, for stars hotter than $5500$\,K. The lower panel shows the fractional difference; the standard deviation of the \textsc{isoclassify}--CHARA difference, $\sigma_{\rm diff}$, is indicated for each mode. 
% Direct mode exhibits a systematic bias that grows with increasing temperature, while grid mode remains consistent with the interferometric scale to within typical A-star temperature uncertainties.
}
\label{f:teff_compare}
\end{figure}

To verify that grid mode is suitable for our sample, we compared its output against interferometric effective temperatures. 
% Interferometry provides a near-direct measurement of \teff\ through the relation
% \begin{equation} \label{eq:interf_teff} F_{\rm bol} \;=\; \frac{1}{4}\,\theta_{\rm LD}^{2}\,\sigma_{\rm SB}\,\teff^{4}, \end{equation}
% where $\theta_{\rm LD}$ is the limb-darkened angular diameter, $F_{\rm bol}$ is the bolometric flux at Earth, and $\sigma_{\rm SB}$ is the Stefan--Boltzmann constant. Provided $F_{\rm bol}$ and $\theta_{\rm LD}$ are measured independently, $\teff$ follows from a direct physical relation, with  weak model dependence entering through the limb-darkening correction and construction of the bolometric flux.
We assembled a comparison sample of 60 stars hotter than $5500$\,K that have published interferometric angular diameters from the CHARA array \citep{Boyajian2012Stellar, Boyajian2012Stellara, Boyajian2013Stellar, White2013Interferometric, White2018Interferometric}, restricted to single, non-variable stars with \textit{Gaia} DR3 counterparts. We used the published $F_{\rm bol}$ and $\theta_{\rm LD}$ values to calculate interferometric temperatures. 

Figure~\ref{f:teff_compare} compares the \textsc{isoclassify} direct- and grid-mode temperatures against the interferometric values. Direct mode shows a clear systematic deviation that grows with increasing temperature, with a standard deviation of $477$\,K about the one-to-one line and a tendency to under-predict \teff\ for the hottest stars in the sample, with fractional discrepancies exceeding $\sim$40\%. This is the expected consequence of the aforementioned diminishing color--temperature sensitivity, confirming that direct mode is unsuitable for this sample. Grid mode, by contrast, shows no significant temperature trend and a standard deviation of $274$\,K, comparable to typical A-type spectroscopic temperature uncertainties \citep{Murphy2019Gaiaderived}. We therefore adopt grid-mode parameters throughout the remainder of this paper.

As a separate check of the grid-mode radius scale, we converted the \textsc{isoclassify}-derived radius $R$ and \textit{Gaia} distance $d$ into an implied angular diameter,
\begin{equation} \label{eq:theta_iso} \theta_{\rm iso} \;=\; \frac{2R}{d}, \end{equation}
which we compared with the measured $\theta_{\rm LD}$. We found good agreement between $\theta_{\rm LD}$ and $\theta_{\rm iso}$ for the grid-mode sample, with a fractional scatter of $\sigma_{\theta}/\theta \approx 4.4\%$, consistent with the combined uncertainties. This angular-diameter comparison tests the inferred radius scale; because $R$ was obtained from the grid-mode fit, it is distinct from the comparison with the independently published interferometric temperatures and is not itself a model-independent temperature validation.

\subsection{Grid-mode parameters for the full sample}
\label{ssec:luminosity}

For each target in our A/F sample we ran \textsc{isoclassify} in grid mode using the MIST isochrones \citep{Choi2016MESA, Dotter2016MESA}. We used \textit{Gaia} DR3 for parallax, $G$, $G_{\rm BP}$, and $G_{\rm RP}$ magnitudes, and the TIC \teff\ as a loose prior with an inflated uncertainty of $\sigma_{\teff} = 300$\,K to avoid over-weighting the photometric TIC temperatures. Extinctions are computed self-consistently within \textsc{isoclassify} using the \texttt{combined19} dust map from the \textsc{mwdust} package \citep{Bovy2016Galactic, Bovy2016Stellar}, which combines the \citet{Marshall2006Modelling}, \citet{Green2015Threedimensional}, and \citet{Drimmel2003Threedimensional} maps into a single all-sky three-dimensional reddening model. The output of each fit is a posterior distribution over \teff, $\logg$, mass, radius, luminosity, and age, from which we adopted the median values as the point estimates and the 16th--84th percentile ranges as the uncertainties. We inflated the uncertainties of the temperature to be in agreement with the expected uncertainty from interferometry ($\sim 300$ K). These parameters form the basis of the analysis presented in the remainder of this paper\rev{, and we list them in Table~\ref{tab:params}}.

\begin{table*}[t!]
\centering
\caption{\rev{Stellar parameters and classifier scores for a subset of the A/F sample. The full table covers all 1{,}546{,}536 targets that we classified. The full table (including all columns) is available in machine-readable form online.}}
\label{tab:params}

\rev{\begin{tabular}{lccccccccc}
\toprule
TIC & $T_{\rm eff}$ (K) & $\log g$ (dex) & $\mathrm{[Fe/H]}$ (dex) & $M$ ($M_\odot$) & $R$ ($R_\odot$) & $\rho$ ($\rho_\odot$) & $L$ ($L_\odot$) & $P(\delta\,{\rm Sct})$ & $\nu_{\rm dom}$ (\cd) \\
\midrule
5751206 & $6610^{+120}_{-130}$ & $3.73 \pm 0.05$ & $0.08^{+0.23}_{-0.21}$ & $1.79^{+0.14}_{-0.12}$ & $3.03^{+0.20}_{-0.19}$ & $0.064^{+0.012}_{-0.011}$ & $15.7^{+3.4}_{-3.1}$ & 0.083 & 23.9651 \\
63888269 & $7660^{+40}_{-80}$ & $4.03 \pm 0.08$ & $-0.19^{+0.32}_{-0.41}$ & $1.65^{+0.16}_{-0.20}$ & $2.04^{+0.18}_{-0.16}$ & $0.188^{+0.052}_{-0.045}$ & $12.9^{+2.3}_{-1.9}$ & 0.936 & 29.9123 \\
237537107 & $7100 \pm 90$ & $3.93 \pm 0.07$ & $-0.08^{+0.30}_{-0.41}$ & $1.63^{+0.15}_{-0.17}$ & $2.28 \pm 0.18$ & $0.135^{+0.032}_{-0.029}$ & $11.9^{+2.2}_{-1.9}$ & 0.926 & 15.3017 \\
275489838 & $6410 \pm 70$ & $4.22^{+0.05}_{-0.06}$ & $0.05^{+0.31}_{-0.27}$ & $1.30^{+0.09}_{-0.10}$ & $1.46^{+0.10}_{-0.06}$ & $0.416^{+0.054}_{-0.081}$ & $3.3^{+0.5}_{-0.4}$ & 0.096 & 6.6271 \\
 & & & & \vdots \\
303728651 & $7710^{+80}_{-60}$ & $3.91^{+0.07}_{-0.06}$ & $0.03^{+0.31}_{-0.32}$ & $1.94^{+0.16}_{-0.18}$ & $2.54^{+0.20}_{-0.19}$ & $0.116^{+0.026}_{-0.024}$ & $20.6 \pm 3.3$ & 0.935 & 15.8246 \\
395001323 & $6470 \pm 90$ & $4.20 \pm 0.06$ & $0.07^{+0.28}_{-0.32}$ & $1.31 \pm 0.10$ & $1.51^{+0.08}_{-0.09}$ & $0.380^{+0.078}_{-0.063}$ & $3.6^{+0.5}_{-0.6}$ & 0.099 & 5.7412 \\
412364990 & $6770 \pm 60$ & $3.97^{+0.04}_{-0.06}$ & $-0.10^{+0.24}_{-0.34}$ & $1.47^{+0.14}_{-0.13}$ & $2.07^{+0.10}_{-0.09}$ & $0.165 \pm 0.025$ & $8.1 \pm 0.9$ & 0.932 & 16.0427 \\
\bottomrule
\end{tabular}}

\end{table*}

% ===============================================================
\section{Identifying $\delta$ Scuti pulsators}
\label{sec:pulsators}

\begin{figure}[t]
\includegraphics[width=\linewidth]{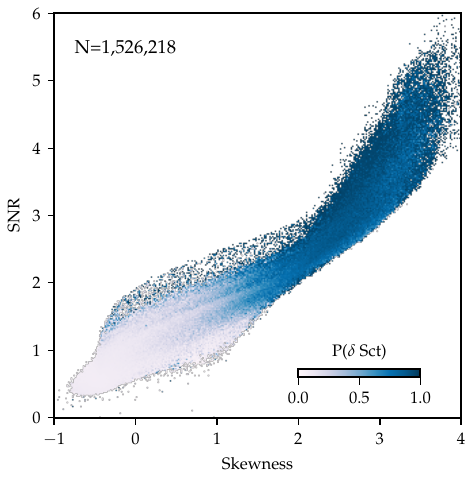}
\caption{Distribution of targets in the spectral skew--SNR plane, colored by the score assigned by the PU classifier. Confirmed \dsct\ pulsators occupy a well-defined locus at high skew and high SNR, separated from the bulk of unlabelled targets clustered at low values of both quantities.}
\label{f:skew_class}
\end{figure}

Our goal is to identify \dsct\ pulsators, in order to map the pulsator distribution across the instability strip and to compute the pulsator fraction. In contrast to the multi-class classification scheme of \citet{Hey2024Confronting}, which simultaneously identified $\gamma$\,Dor, eclipsing, and other OBAF variables using a random forest, here we attempt only to classify whether a target shows \dsct-like pulsation or not.

A challenge with this classification is the absence of a curated negative class. Catalogues of manually vetted known \dsct\ pulsators are extensive \citep{Zhou2023Catalog, Zhou2025Identification}, but no comparably comprehensive list of confirmed {non-pulsators} within the instability strip exists, beyond extrapolation of \citet{Murphy2019Gaiaderived}. Any star not labelled as a \dsct\ in the literature could equally be a genuine non-pulsator, an undetected low-amplitude pulsator, or simply a target that has never been inspected. Standard supervised classification assumes both positive and negative training labels are reliable, and treating unlabelled stars as negatives in this setting biases the resulting classifier toward under-predicting the positive class.

\subsection{Positive--Unlabelled classification}
\label{ssec:pu_method}

We address this with a Positive--Unlabelled (PU) learning approach, recently introduced to asteroseismic classification by \citet{Kliapets2025Automated}. In the PU framework, a labelled set of confirmed positives is combined with a large unlabelled set containing an unknown mixture of true positives and true negatives. The classifier can then rank candidates without requiring every unlabelled target to be treated as a secure negative. This makes PU learning well suited to our problem, where the positive class (\dsct\ variables) is well characterised but the unlabelled population mixes genuine non-pulsators with currently unclassified pulsators.

We classified these stars with the bagging PU algorithm of \citet{mordelet14}, using a support
vector machine \citep[SVM;][]{cortes95} from \texttt{scikit-learn}
\citep{scikit-learn} as the base estimator. 
% The algorithm builds an ensemble of
% $K$ binary classifiers. 
% Each classifier receives every labelled positive
% together with a bootstrap sample of the unlabelled stars, and it is trained to
% separate the two groups as though the unlabelled stars were all negative. The
% pulsators hidden in each bootstrap sample act as label noise. Each classifier
% sees a different random draw of that noise, so its effect is reduced when the
% ensemble outputs are averaged, and the averaged score ranks every star by its
% similarity to the labelled pulsators.

\subsection{Features}
\label{ssec:features}

For each target we computed a small set of features summarising the Fourier spectrum and light-curve morphology, loosely based on those used by \citet{Hey2024Confronting}. We used (i)~the amplitude and frequency of the dominant Fourier peak above $5$\,\cd; (ii)~the signal-to-noise ratio (SNR) of that peak relative to the $95$th-percentile noise floor; (iii)~the number of significant, independent peaks exceeding a false-alarm level of 1\%; (iv)~the skewness of the amplitude spectrum above $5$\,\cd; and (v)~the skewness of the light curve. The first two capture the presence and location of coherent power in the \dsct\ range, the third captures the multiperiodic character of the class, and the last two discriminate between the heavy-tailed amplitude distribution of a pulsator and the approximately symmetric distribution of a pure-noise spectrum.

Figure~\ref{f:skew_class} illustrates the separability of the \dsct\ class in the spectral-skew--SNR plane, which is a typical diagnostic (i.e., \citealt{Murphy2019Gaiaderived, Dholakia2024Catalog, Read2024Identifying}). Confirmed pulsators occupy a well-defined region at high values of both quantities, while the bulk of unlabelled targets cluster at low skew and low SNR. The two populations overlap in a transition region that includes low-amplitude pulsators, where the PU classifier relies on the remaining features to make a decision.

\subsection{Training and validation}
\label{ssec:training}

The positive training set was taken from the manually-vetted \tess\ \dsct\ catalogues of \citet{Zhou2023Catalog, Zhou2025Identification}, which together provide $\sim$$100{,}000$ confirmed pulsators covering the full sky within our adopted temperature range. To assess the purity of this catalogue, we randomly drew $1{,}000$ targets, retrieved their \textsc{TGLC} light curves, and independently re-classified them by visual inspection of the light curve and amplitude spectrum. We found an agreement of $96\%$, with the small residual disagreement dominated by what appear to be borderline low-amplitude cases. We retained the full positive sample, noting that this residual contamination is small enough that its effect on the trained classifier is negligible compared to the contamination of the unlabelled set by undetected positives.

We held back $20\%$ of the positive set as a validation sample, spanning the full range of apparent magnitude and effective temperature to ensure representative coverage of the feature space. We trained the classifier on the remaining $80\%$, together with the full unlabelled set. The final classifier achieved a recovery rate of {$95\%$} on the held-back positives with classification labels exceeding 0.9. However we point out that because the classifier is trained against unlabelled stars, $p_{\delta\,{\rm Sct}}$ is a ranking score. It is not a calibrated probability that a star is a $\delta$~Scuti pulsator, and we therefore choose the classification threshold empirically. Applying the classifier to the full sample of \samplesize\ stars and adopting a classifier threshold of {$P(\dsct) > 0.9$} yielded \ndsct\ \dsct\ pulsators. Our threshold is based largely on the recovery of the held-back positives.

The PU classifier identifies any star whose periodogram is dominated by sharp, high-SNR Fourier peaks in the \dsct\ frequency range, which can include sources of variability other than self-excited pressure modes. We therefore applied an additional cut. We removed known eclipsing binaries and ellipsoidal variables, whose orbital harmonics can populate the \dsct\ frequency range and mimic pulsation features in the Fourier spectrum \citep{IJspeert2024Statistical}. To do this, we cross-matched our sample with the \tess\ Eclipsing Binary catalogue \citep{Prsa2022TESS} and the \textit{Gaia} DR3 eclipsing-binary catalogue \citep{Mowlavi2023Gaia}, removing {$3,908$} targets in total. This cut necessarily removed genuine \dsct\ variables in eclipsing systems; since the goal of this paper is a population-level analysis, and since the pulsation properties of such systems cannot be measured homogeneously alongside the single stars, we accepted this loss. The removed targets are flagged in \rev{Table~\ref{tab:params}}.

% ===============================================================
\section{The H--R diagram and pulsator fraction}
\label{sec:results}

\begin{figure*}[t]
\centering
\includegraphics{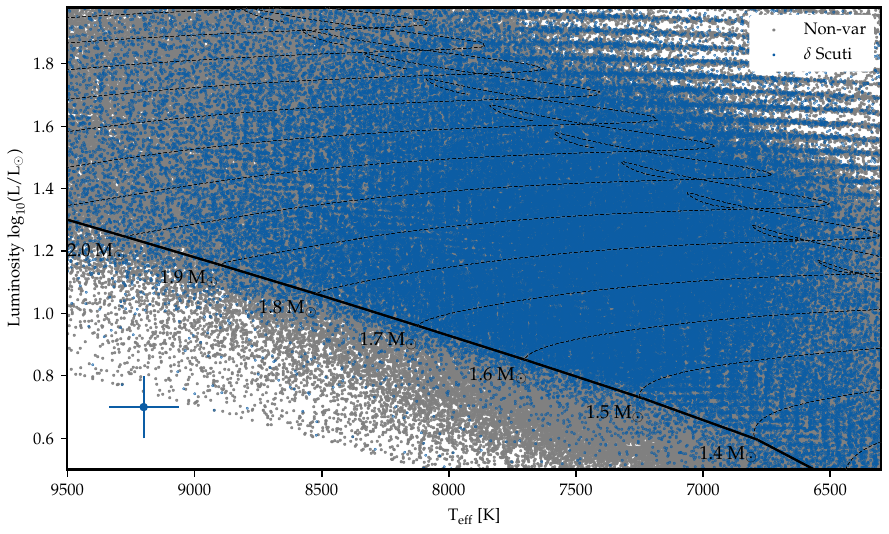}
\includegraphics{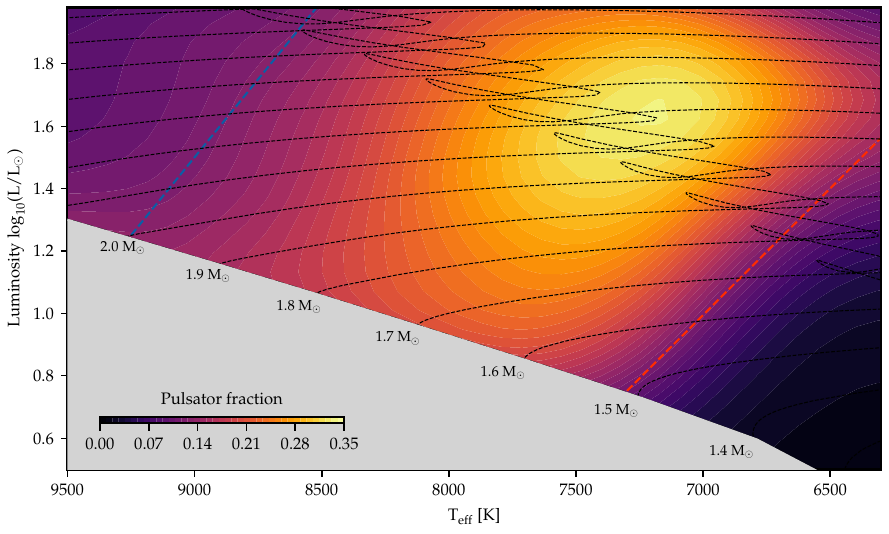}
\caption{\emph{Top:} H--R diagram of the full sample, including stars with classification probability $<0.9$ (gray) and \dsct\ candidates with a classifier score $>0.9$ (blue). MIST evolutionary tracks are labelled by mass. \emph{Bottom:} pulsator fraction in bins of \teff\ and $\log(L/\Lsun)$, with bin boundaries restricted to contain at least $50$ targets each. The peak pulsator fraction is $\sim 30\%$, reached near $\teff \approx {7150}$\,K.}
\label{f:hrd_full}
\end{figure*}

Figure~\ref{f:hrd_full} shows the H--R diagram of the full sample, with classified \dsct\ pulsators highlighted. The lower panel shows the \dsct\ pulsator fraction computed in two-dimensional $(\teff, \log L)$ bins. The pulsators trace a clearly defined region that closely follows the classical instability strip, with a sharp red edge and a more diffuse blue edge. The evolutionary tracks shown there were computed as follows.

We used {\sc mesa} r24.03.1 to calculate tracks with masses from 1.3 to 3.0\,${\rm M}_\odot$ in steps of 0.1\,${\rm M}_\odot$ at $Z = 0.0142$, corresponding to the bulk solar metallicity \citep{Asplund2009Chemical}, using the input physics and inlist template of \citet{gautam2026grid}. We included rotation, from the ZAMS onwards, by first evolving the tracks without rotation from the pre-main sequence to 50\,Myr. Their zero-age main sequence (ZAMS) ages were then located on that track with a variant of the stabilisation criterion of \citet{gautam2026grid}. Specifically, for each of the effective temperature and the mean density we take the rolling standard deviation, over 100 consecutive models, of its smoothed second derivative with respect to age, and divide it by that quantity's threshold. The ZAMS is the first age at which the mean of these two ratios falls below one. This differs from the criterion as described by \citet{gautam2026grid} in dropping its $\Delta \nu$ term and in averaging the terms rather than requiring each to lie below its threshold. The model nearest the ZAMS is loaded into a new run, in which we switched on \textsc{mesa}'s rotation and relaxed the star to uniform rotation at $\Omega = 0.4\,\Omega_{\rm crit}$. After relaxation, a uniform angular-momentum diffusion coefficient with \textsc{mesa}'s default value, \texttt{uniform\_am\_nu\_non\_rot} $= 10^{20}\,{\rm cm^2\,s^{-1}}$, was used to set a solid-body rotation profile, and $\Omega/\Omega_{\rm crit}$ is left to change as the star evolves. Neither rotational mixing of chemical elements nor mass loss was included. During the rotating evolution, \textsc{mesa} uses adaptive timesteps with a maximum of $2\,\mathrm{Myr}$ (\texttt{max\_years\_for\_timestep} $= 2\times10^6$). We retained additional timestep controls using in $\log_{10} T_{\rm eff}$ and $\log_{10}(L/L_\odot)$, with \texttt{delta\_lgTeff\_limit} $= 0.0006$ and \texttt{delta\_lgL\_limit} $= 0.002$. 

A \textsc{mesa} profile and its pulsation input file in the GSM format of \textsc{gyre} were written at every fifth model and at the model on which the run stopped. For every profile, with no cut on evolutionary phase, we used \textsc{storm} (\citet{Vanlaer2026VincentVanlaer}) to compute mode frequencies for degrees $\ell \leq 3$, accounting for (uniform) rotation at the model's surface angular velocity $\Omega$. We scanned 500 trial frequencies for each $(\ell, m)$, across a scan window of 0.25--150\,d$^{-1}$. We ultimately used just the radial mode frequencies for this work.

\subsection{The observational instability strip}
\label{ssec:edges}

\begin{figure*}[t]
\centering
\includegraphics[width=\linewidth]{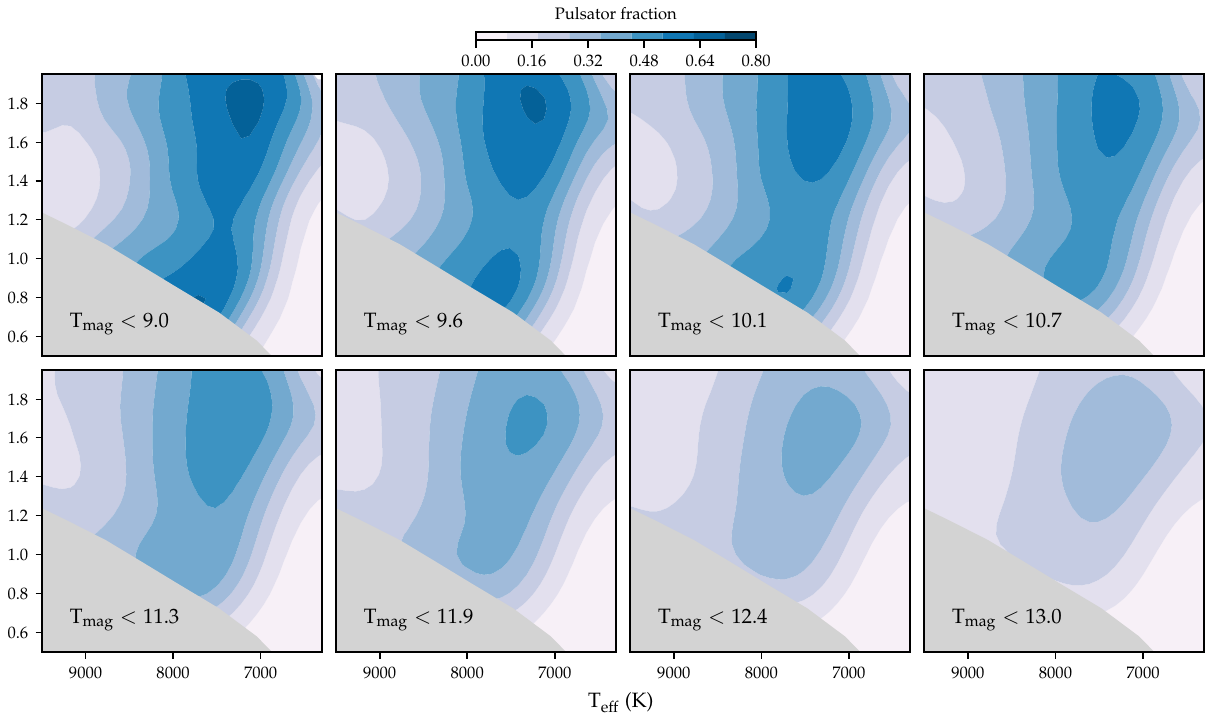}
\caption{The pulsator fraction across the $\teff$--$\log(L/\Lsun)$ plane, computed for progressively fainter magnitude-limited subsamples. The peak fraction falls monotonically from $\sim {80}\%$ for $T_{\rm mag} < 9.0$ to $\sim {20}\%$ for $T_{\rm mag} < 13.0$.
% demonstrating that a substantial part of the apparent impurity of the instability strip in the full sample is due to reduced sensitivity at fainter magnitudes.
}
\label{f:pfrac_tmag}
\end{figure*}

To measure the edges of the observational instability strip, we require a smooth estimate of the pulsator fraction across the $\teff$--$\log L$ plane. To construct this estimate we made use of a simple Gaussian convolution, which we now detail. We constructed bins of width (1.0, 2.0) in $\teff$--$\log L$. Within each bin, $(a,b)$, we counted the total number of stars $n_{ab}$ and the number of $\delta$~Scuti pulsators $k_{ab}$, giving a fraction of $f_{ab} = k_{ab}/n_{ab}$. This estimator is binomial and its variance, $f(1-f)/n$, is dominated by the sparsely populated bins at the edges of the strip, producing jagged contours that preclude a direct edge fit. Rather than smoothing $f_{ab}$ directly, which would equally weight small and large numbered bins, we convolved the count fields separately and took their ratio,
\begin{equation}
\label{eq:smoothed_fraction}
\hat{f} = \frac{G_{\boldsymbol{\sigma}} * k}{G_{\boldsymbol{\sigma}} * n},
\end{equation}
where $G_{\boldsymbol{\sigma}}$ is a Gaussian kernel and $*$ denotes convolution. This ensured that under-populated bins contribute negligibly to the contour. We adopted an anisotropic kernel, $\boldsymbol{\sigma} = (\sigma_{\log L}, \sigma_{\teff})$ with $\sigma_{\log L} > \sigma_{\teff}$. Because the strip edges are steep in temperature but not with increasing luminosity, smoothing more in the luminosity direction suppresses bin-to-bin noise while preserving the sharp temperature gradient that defines the edge, and adopt a $\sigma=(1.0, 2.0)$.

The effective number of stars contributing to each smoothed pixel is
\begin{equation}
\label{eq:neff}
N_{\mathrm{eff}} = \frac{\left(\sum_{ab} w_{ab}\, n_{ab}\right)^2}{\sum_{ab} w_{ab}^2\, n_{ab}}.
\end{equation}
Here, $w_{ab}$ are the kernel weights, which we use both to propagate a per-pixel uncertainty, $\sigma_{\hat{f}} = [\hat{f}(1-\hat{f})/N_{\mathrm{eff}}]^{1/2}$, and to mask regions with $N_{\mathrm{eff}} < N_{\mathrm{min}}$ where the fraction is too poorly constrained to trace an edge. 

We defined the observational instability strip from this smoothed pulsator-fraction field, $\hat{f}$. Rather than adopting the fixed 20\% contour used by \citet{Murphy2019Gaiaderived}, we chose the contour that encloses {$80\%$} of the detected pulsators, weighting each bin by its unsmoothed pulsator count. In each luminosity row, we linearly interpolated the red and blue crossings surrounding the local maximum in $\hat{f}$ and fitted linear relations to obtain
\begin{align}
    \log_{10}(L)_{\rm red} &= -0.000669 (\pm0.0002) * {\rm T}_{\rm eff} + 5.724 (\pm 0.01) \\ 
    \log_{10}(L)_{\rm blue} &= -0.001030 (\pm 0.0004) * {\rm T}_{\rm eff} + 10.59 (\pm 0.02).
\end{align}
% \begin{equation}
% T_{\rm edge}=A_{\rm edge}
% +B_{\rm edge}\left[\log(L/\Lsun)-\log(L/\Lsun)_0\right].
% \end{equation}
The strip between these lines contains {$80\%$} of the detected pulsators for the entire sample. 
% We do seek to point out, however, that the observational instability strip appears to be a non-linear function of temperature and luminosity, such that it would be better represented as, perhaps, a gaussian blob or some such.

We estimated the edge uncertainties with {$1000$} bootstrap realizations. In each step, we resampled the stars with replacement, perturbed their effective temperatures and luminosities according to their uncertainties, reconstructed the smoothed fraction map, and repeated the contour selection and edge fit. We adopted the median bootstrap edge as the fitted boundary and the 16th and 84th percentiles as its 68\% confidence interval.

% We then fit the blue and red edges as straight lines in the $\teff$--$\log L$ plane, taking the edge location in each luminosity row to be the temperature at which $\hat{f}$ crosses a fixed threshold of \todo{$X$}. We verified that the inferred edge locations are insensitive to the kernel width across a range of $\sigma_{\teff}$, confirming that the smoothing does not bias the measured strip boundaries; the residual drift over this range is propagated into the systematic error budget on the edge positions. The resulting edges are
% \begin{equation}
% \label{eq:edges}
% \log(L/\Lsun) = \todo{m_{\rm blue}}\,\teff + \todo{c_{\rm blue}}, \quad
% \log(L/\Lsun) = \todo{m_{\rm red}}\,\teff + \todo{c_{\rm red}}.
% \end{equation}

\begin{figure*}[t]
\centering
\includegraphics[width=0.39\linewidth]{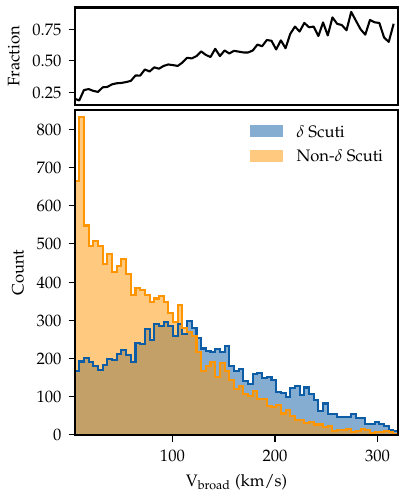}
\includegraphics[width=0.6\linewidth]{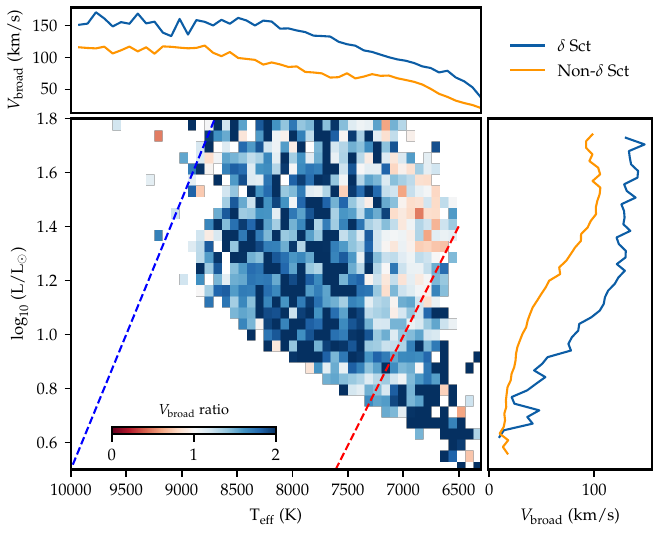}
\caption{ \emph{Left:} distribution of \vbroad\ for the two populations within the instability strip, with the pulsation fraction in the top panel. \emph{Right:} the ratio of the median \textit{Gaia} \vbroad\ of classified \dsct\ pulsators to that of non-pulsating stars, across the $\teff$--$\log(L/\Lsun)$ plane, for targets brighter than $T_{\rm mag} = 13$ and bins containing at least 5 \dsct\ variables. Values above unity (blue) indicate that pulsators rotate faster than non-pulsators at the same position in the H--R diagram. The dashed lines mark the derived instability strip from Sec.~\ref{ssec:edges}. The marginal panels show the median \vbroad\ of each population as a function of \teff\ (top) and $\log L$ (right).}
\label{f:vbroad}
\end{figure*}

This full-sample calculation is a lower limit on the intrinsic pulsator fraction, because our sensitivity to low-amplitude pulsation degrades towards fainter magnitudes (see \citealt{Berry2026Occurrence}). Figure~\ref{f:pfrac_tmag} shows the pulsator fraction recomputed for a series of magnitude-limited subsamples. The peak fraction rises monotonically with brightness, from {$\sim 20\%$} for the full $T_{\rm mag} < 13.5$ sample to {$\sim 80\%$} for $T_{\rm mag} < 9.0$. {The same trend was found by \citet{Read2024Identifying} in a color-restricted subsample of the same 10-min FFI data; they reported a pulsator fraction of ${\sim}70\%$ at $G<8$ falling to ${\sim}45\%$ at fainter magnitudes. Subsequently, \citet{Mani2025Characterizing} found a pulsator fraction of ${\sim}70\%$ for bright ($G<7$) stars within one magnitude of the ZAMS. Our all-sky sample extends this behavior across the full strip and over a wider magnitude range.} \rev{Crowding has a similar effect. At fixed magnitude, the noise level of the light curves increases towards the Galactic plane, and the fraction of instability-strip stars that we detect as pulsators falls accordingly (Appendix~\ref{app:blending}).}

The general morphology of the strip on the red end is preserved across all magnitude bins in Figure~\ref{f:pfrac_tmag}. That is, the peak remains at the same location in $(\teff, \log L)$ and the edges do not move appreciably, but its amplitude scales with detectability.
This implies that the observational instability strip is considerably more populated than the full-sample fraction suggests, and a large part of the difference between our ${\sim}30\%$ and the ${\sim}70\%$ of \citet{balonaKeplerObservationsScuti2011} and \citet{Murphy2019Gaiaderived} is attributable to the shallower effective sensitivity of \tess\ FFI photometry relative to \kepler\ {\citep[cf.][who show that a single \tess\ sector reaches only ${\sim}50$\,ppm at $G \approx 9$]{Read2024Identifying}}. Second, even in the brightest bin the peak fraction does not approach unity; {$\sim 20\%$} of stars at the centre of the strip remain undetected as pulsators. The strip is therefore not pure, but is more pure than the full-sample number would imply, and our result is consistent with the \kepler-based determinations.

\subsection{The influence of rotation}
\label{ssec:rotation}

Recent work has highlighted the role of rotation in shaping the \dsct\ population \citep{Gootkin2024New, Wang2025Rotation, Murphy2026Population}. Our larger sample, combined with the better frequency coverage of the 10-minute cadence, allows us to revisit this question with improved statistics.

Figure~\ref{f:vbroad} shows the distribution of the \textit{Gaia} \vbroad\ line-broadening parameter for classified \dsct\ pulsators and for non-pulsating stars within the instability strip, restricted to the subset of targets with \vbroad\ measurements. The left panel shows that \dsct\ variables are, on average, faster rotators than their non-pulsating counterparts. The median \vbroad\ of \dsct\ pulsators is {$114$\,km\,s$^{-1}$}, compared with {$68$\,km\,s$^{-1}$} for the non-pulsators, and a two-sample Kolmogorov--Smirnov test rejects the null hypothesis that they are drawn from the same distribution. Critically, the right panel of Figure~\ref{f:vbroad} shows that this offset is not an artefact of the two populations occupying different regions of the H--R diagram; the ratio of median \vbroad\ exceeds unity across essentially the whole strip, so the effect holds at fixed \teff\ and $\log L$. 

The pulsator fraction increases monotonically with \vbroad\ up to {$\sim 200$\,km\,s$^{-1}$} and plateaus at {$\sim 75\%$} for higher rotational velocities (Fig.~\ref{f:vbroad}), in agreement with \citet{Gootkin2024New} and consistent with the expectation that rapid rotation inhibits the gravitational settling of helium out of the \ionn{He}{ii} ionisation zone, sustaining the $\kappa$ mechanism \citep{Theado2011Diffusioninduced, Theado2005Helium, Theado2003Coupling, Theado2003Couplinga}.

The existence of slowly rotating \dsct\ pulsators, however, is not naturally accommodated by this picture. We find a pulsator fraction of {$\sim 25\%$} for stars with $\vbroad < 50$\,km\,s$^{-1}$, well within the regime where Am stars dominate. Some fraction of these will be rapid rotators seen close to pole-on, for which $\vbroad \simeq v\sin i$ is small despite a large equatorial velocity. For an isotropic distribution of inclinations, however, only a small percentage of stars with high equatorial velocities would appear at low \vbroad\ due to their inclination, which is insufficient to account for the observstions. The metallic-line character of these slowly rotating pulsators, and the corresponding helium abundance in their envelopes, will require spectroscopic follow-up before this population can be physically interpreted.

\subsection{Variability characteristics across the strip}
\label{ssec:freq_hrd}

The 10-minute \tess\ cadence allows us to probe the \dsct\ pulsation spectrum with a Nyquist frequency of $72$\,\cd, in contrast to the \kepler\ long-cadence ($24$\,\cd\ Nyquist) and \tess\ 30-minute FFI ($24$\,\cd\ Nyquist) datasets used in previous large-sample studies \citep{Murphy2019Gaiaderived, Gootkin2024New}. In those data, frequencies above the nominal Nyquist limit can appear as reflected aliases, and their observed amplitudes are reduced by smearing from the nonzero integration time (apodization; \citealt{Murphy2013SuperNyquist, Shibahashi2018SuperNyquist, Murphy2015Potential}). Although individual super-Nyquist modes can sometimes be recovered, attenuation complicates population-level selection. The 10-minute data move the Nyquist frequency to $72$\,\cd\ and substantially reduce this problem, but do not eliminate it. Although modes above $72$\,\cd\ remain aliased, we treat frequencies approaching the Nyquist limit as unaliased but amplitude-attenuated measurements \cite{2026A&A...710A.245M}.

\begin{figure}[t!]
\centering
\includegraphics[width=\linewidth]{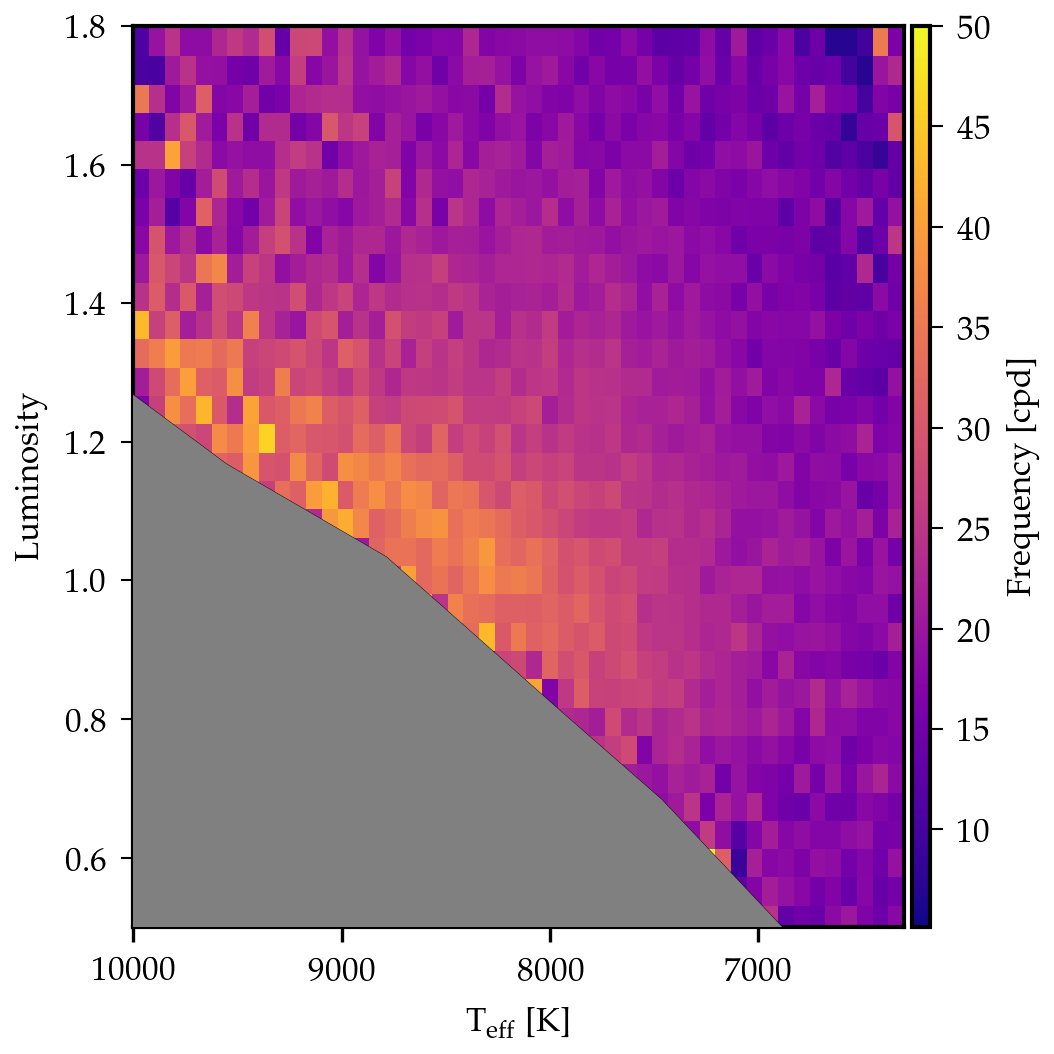}
\caption{H--R diagram of classified \dsct\ pulsators, color-coded by the frequency of the dominant Fourier peak. Cooler, less luminous pulsators near the red edge oscillate at lower frequencies, while hotter, more luminous pulsators near the blue edge exhibit higher dominant frequencies, consistent with the expected scaling with mean stellar density.}
\label{f:freq_hrd}
\end{figure}

Figure~\ref{f:freq_hrd} shows the H--R diagram of pulsators color-coded by their dominant pulsation frequency. A clear gradient is visible across the instability strip, with cooler, less luminous pulsators near the red edge exhibiting dominant frequencies of {$\sim 10$--$20$\,\cd}, while hotter, more evolved pulsators near the blue edge cluster at {$\sim 50$--$70$\,\cd}. This trend is primarily the expected consequence of the scaling of acoustic mode frequencies with the square root of the mean stellar density \citep{Aerts2010Asteroseismology, Breger2000Scuti}: as stars evolve off the zero-age main sequence their radii increase and their mean density falls, lowering the frequency of any given radial order.

\subsection{Prewhitening the full sample}
\label{ssec:prewhiten}

\begin{figure}[t!]
\centering
\includegraphics[width=\linewidth]{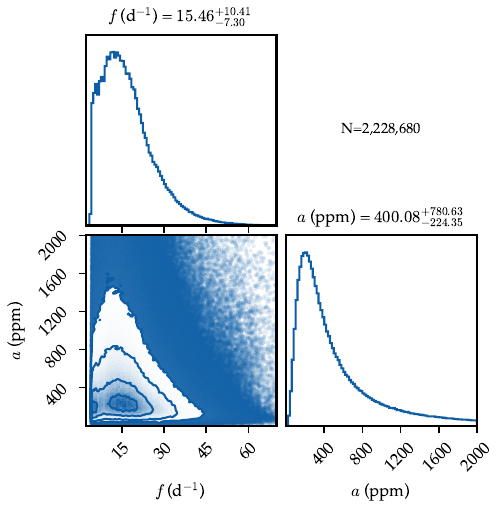}
\caption{Joint distribution of extracted frequency $f$ and amplitude $a$ for all significant peaks prewhitened from the \dsct\ sample with classifier score $\geq 0.9$. The marginal distributions and their medians with 16th--84th percentile ranges are shown in the upper and right panels.}
\label{fig:prew_all}
\end{figure}

To characterize the pulsation spectra beyond the dominant mode, we iteratively prewhitened each light curve of candidates with classifier score $\geq 0.9$. At each iteration we identified the highest peak in the amplitude spectrum, optimized its frequency, amplitude, and phase by non-linear least squares against the residual light curve, subtracted the resulting sinusoid, and repeated until no peak exceeded {a false-alarm probability of 1\%}. This process was identical to that of \cite{Hey2024Confronting}. Peaks separated by less than {$10^{-2}$\,\cd} from a stronger peak were treated as duplicates and removed. This process yielded $2{,}228{,}680$ significant frequencies across our \dsct\ sample.
% \footnote{We make available our complete prewhitened results in machine-readable format online \todo{ZENODO}}

The resulting joint distribution of frequency and amplitude is shown in Figure~\ref{fig:prew_all}. The frequency distribution peaks at $f = 15.48$\,\cd\ and is strongly right-skewed, with a tail extending to the Nyquist limit; the amplitude distribution is approximately log-normal, centred on a median amplitude of {$\sim 400$\,ppm}. There is a weak correlation between frequency and amplitude across the population, with increasing frequencies corresponding to lower amplitudes, indicating that the amplitude distribution of excited modes is weakly dependent of where in the \dsct\ frequency range they lie \citep[e.g.,][]{Balona2018Gaia, Bowman2018Characterizing, Mani2025Characterizing}.

% ===============================================================
\section{The period--luminosity and period--density relations}

\label{sec:pl}

\begin{figure*}[t!]
\centering
\includegraphics[width=0.99\linewidth]{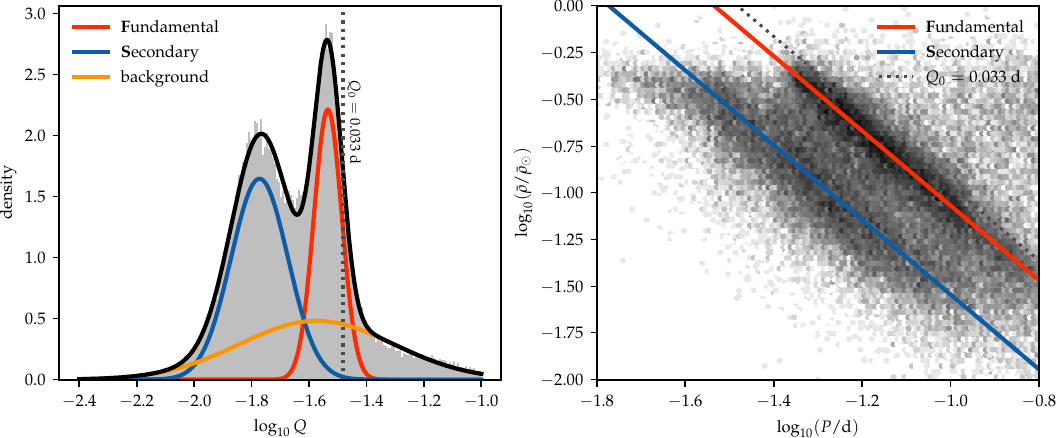}
\caption{\textit{Left}: The representative pulsation constant $Q$ of the sample following Eq.~\ref{eq:q}, fit with three Gaussian distributions corresponding to the fundamental (red), secondary (blue), and background ridges. The dashed black line marks the canonical fundamental mode pulsation constant \citep{Breger2000Scuti}. \textit{Right}: The period--density relation with the straight lines obtained from the Gaussian means to the pulsation constant.}
\label{f:prho}
\end{figure*}

The \dsct\ variables, like other types of p-mode pulsators, exhibit a period--luminosity (PL) relation \citep{McNamara2011Deltaa}. This arises from the more fundamental relation between period and density \citep[e.g.,][]{catelanPulsatingStars2015}, as characterized by the pulsation constant
\begin{equation}
\label{eq:pulsation_constant}
Q \;=\; P \left( \frac{\bar{\rho}}{\bar{\rho}_{\odot}} \right)^{1/2}.
\end{equation}
Here, $P$ is the pulsation period and $\bar{\rho}$ is the mean stellar density \citep{Breger2000Scuti}. For a given mode, $Q$ is approximately constant across the \dsct\ class. The period of this mode is set by the star's mean density, which explains the period--density relation. On the main sequence, density scales approximately with luminosity, which explains the existence of the PL relation. Here we present both, but emphasise the period--density relation \citep[as in][]{Mani2025Characterizing}, which links directly to Equation~\ref{eq:pulsation_constant}.

For all of the fits we made the following cuts on our data. We selected stars with classifier probability $p_{\rm class}\geq0.90$, dominant-peak $\mathrm{SNR}\geq10$, and catalogue spectral-skewness statistic $\mathtt{logskewa}\geq1.5$. We additionally required a positive parallax with $\sigma_\varpi/\varpi<0.05$, $\mathrm{RUWE}<1.4$, $-1.8\leq\log_{10}(P_{\rm dom}/{\rm d})\leq-0.8$, $\sigma_{\log L}\leq0.15$\,dex, and $\sigma_{M_G}\leq0.30$\,mag. For the period--luminosity fit, we also required that the adopted extinction satisfy $0\leq A_V\leq5$\,mag and the inferred intrinsic color $-0.2\leq(G_{\rm BP}-G_{\rm RP})_0\leq1.5$. These cuts left $93{,}971$ stars.

\subsection{Period--density relation}
\label{sec:pdensity}
We begin with the period--density fit, shown in Fig.~\ref{f:prho}. Assuming constant-$Q$ sequences, we define
\begin{align}
    x &\equiv \log_{10} \frac{P}{\rev{\mathrm{d}}},\\
    y &\equiv \log_{10} \frac{\rho}{\rho_\odot},
\end{align}
such that
\begin{equation}
    y = -2x + 2\log_{10} Q.
    \label{eq:fixed}
\end{equation}

Recasting the period--density relation to this form is useful for recovering the pulsation constants directly (as in \citealt{Murphy2026Population}). Rather than fitting arbitrary straight lines in the $(x,y)$ plane, we instead impose a fixed slope from Eq.~\ref{eq:fixed} and project each star into the one-dimensional coordinate
\begin{equation}
    q \equiv x + \frac{1}{2}y = \log_{10}\frac{Q}{\rev{\mathrm{d}}}.
    \label{eq:q}
\end{equation}
Note that this is equivalent, within a multiplicative constant, to measuring the perpendicular distance to a line with slope~$-2$ \citep[see][]{Ziaali2019Period, Barac2022Revisiting, Read2024Identifying, Mani2025Characterizing}.

Every parallel constant-Q ridge becomes a peak in the $q$ distribution, and its mean $\mu$ maps back to a line with intercept 2$\mu$ in Eq.~\ref{eq:fixed}, and as seen in Fig.~\ref{f:prho}. This formulation fit $Q$ directly and reduces the problem to fitting an arbitrary sum of mixed Gaussian distributions, allowing us to easily incorporate background distributions. 

We modeled the distribution of $q$ as a sum of $K$ Gaussian ridge components plus a broad Gaussian background:
\begin{align}
    p(q_i) &= \sum_{k=1}^K\omega_k \mathcal{N}(q_i|\mu_k, s_{i,k}) + \omega_{\rm bg} \mathcal{N}(q_i | \mu_{\rm bg}, s_{i, \rm bg})
\end{align}
where $s_{i,k}$ is the standard deviation characterizing the width of the $k$th ridge. The broad component absorbs non-radial modes, higher radial orders, and contaminants. We maximized the likelihood of this model and compared models with $K=1, 2, 3$ Gaussian mixtures using the Bayesian Information Criterion (BIC; \citealp{Schwarz1978Estimating}). A single ridge is strongly disfavoured, as is plainly obvious from Fig.~\ref{f:prho} ($\Delta BIC=2401$), while the two- and three-ridge models have similar BIC values ($-49524$ and $-49589$, respectively). For simplicity, we adopted the two-ridge model, while noting that it is possibly a combination of several higher-order overtones (i.e., 3O + 4O) and unresolved binaries, which we discuss further below. We label the high-$Q$ sequence as the fundamental mode ($F$) and the lower-$Q$ sequence as a secondary ridge ($S$).

For the two-ridge mixture, we obtained $Q_F = 0.029216 \pm 0.0001$ and $Q_{S} = 0.016892 \pm 0.0001$, giving corresponding period--density relations of
\begin{align}
    \log_{10}(\rho/\rho_\odot)_{\rm F} &=-2 \log_{10}\Big(\frac{P}{\rev{\mathrm{d}}}\Big) - 3.0616,\nonumber \\
    \log_{10}(\rho/\rho_\odot)_{\rm S} &=-2 \log_{10}\Big(\frac{P}{\rev{\mathrm{d}}}\Big) - 3.6048.
\end{align}
The separation between the ridges is 0.2716 dex in $\log P$ at fixed density, or 0.5432 dex in $\log\rho$ at fixed period, and their ratio is $Q_{\rm S}/Q_{\rm F}=0.578$. The fundamental ridge lies close to, but below the commonly adopted fundamental value of $Q=0.033$ d. This difference could potentially be caused by population variation in $Q$ or a systematic offset in the density scale, which we now discuss.

The second sequence has been previously identified in the \dsct\ period--luminosity diagram by \citet{Ziaali2019Period} and \citet{Barac2022Revisiting}. Taking the high-Q sequence to be the radial fundamental, its value can be compared directly with the $Q_{\rm F}=0.0315$ suggested by \citet{Barac2022Revisiting} from \kepler\ data. Our value is slightly lower but both are within the uncertainty. From the tracks used in Sec.~\ref{sec:pulsators}, we calculate a mass-dependent $Q_F$ between $0.0311$ and $0.0329$, quite far below the canonical value and close to that of \citet{Barac2022Revisiting}. Our value lies slightly below this range, and we note that the PL relation investigated in \citet{Barac2022Revisiting} was performed with \kepler\ data of a much higher photometric quality.

The ratio between the two sequences we measure is {$Q_{\mathrm{S}}/Q_{\mathrm{F}} = 0.578 \pm 0.001$}. The ratio is a useful value to consider here because it is invariant under any rescaling of the density zero point. That is, a systematic error in $\bar{\rho}$ (or period, for that matter) displaces both sequences equally in $\log_{10}Q$ and leaves their ratio unchanged. 
Our ratio is slightly larger than the value reported by \citet{Ziaali2019Period} and \citet{Barac2022Revisiting}, who measured the second ridge to be a factor of two shorter in period. They suggested it might arise from a $2:1$ resonance between the fundamental mode and either the third or fourth overtone. Our ratio is largely consistent with this result. The theoretical values for $Q_2/Q_0 = 0.621$ and $Q_3/Q_0 = 0.515$ straddle our measured ratio \citep{coxTheoreticalPulsationConstants1972}. Likewise, {\citet{Murphy2026Population} recently argued that the second ridge is satisfactorily reproduced with a mixture model consisting of the same 2:1 resonance, and a term representing a combination frequency of the fundamental and first overtone modes.}
%Likewise, assuming an unresolved equal-mass binary forms the secondary ridge gives a ratio of $Q_{\rm binary} / Q_0 = 0.594292$, which is only $0.01$ dex away from our measured ratio.  Without better precision on our measurements we can not accurately attribute the secondary ridge to any single process. However, we note that the much larger intrinsic width of the distribution points to it being a likely combination of several overtone ridges and perhaps unresolved equal-mass binaries.
%\trb{a significant population of equal-mass binaries seems very unlikely, I think this suggestion is probably a distraction}

% \todo{the width of the second ridge is much bigger. Probably a combination of overtones and equal mass binaries}
% We caution that the secondary ridge (lower-$Q$) may not be a distinct sequence at all. Its fitted intrinsic width is approximately three times that of the fundamental sequence. 

\subsection{Period--luminosity relation}

\begin{figure}[t]
    \centering
    \includegraphics[width=\linewidth]{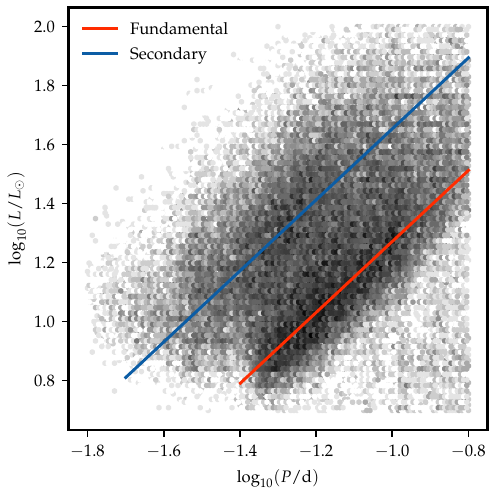}
    \caption{The period luminosity relation and corresponding fits for the fundamental (red) and secondary (blue) ridges.}
    \label{fig:pl}
\end{figure}

The period--luminosity relation for the dominant mode in our sample is shown in Fig.~\ref{fig:pl}, using the luminosities returned by \texttt{isoclassify}. We fitted the ridges in both the \texttt{isoclassify} reported luminosity \rev{and the \textit{Gaia} $G$-band absolute magnitude, $M_G$. We first corrected the apparent $G$-band magnitude for extinction, $G_0 = G - A_G$, and then computed the absolute magnitude} as
\begin{equation}
    \rev{M_{G} = G_0} + 5\log_{10}\!\left(\frac{\varpi}{\mathrm{mas}}\right)-10,
    \label{eq:absmag}
\end{equation}
where $\varpi$ is the \textit{Gaia}-DR3 parallax and $A_G$ is obtained from the line-of-sight $V$-band extinction $A_V$, as returned by the \texttt{isoclassify} posterior. Because the \textit{Gaia} passband is broad, the ratio $A_G/A_V$ is not constant but depends on the intrinsic color (and reddening) of the star. We thus adopted the polynomial coefficients of \citet{Luhman2020} who fitted $A_{G, \rm BP, \rm RP}/A_V$ as functions of the de-reddened Gaia color. We detail the process of obtaining $A_G$ in Appendix~\ref{app:extinction}.

Given our values of temperature, luminosity, and \rev{$M_G$}, we fitted both sequences simultaneously with
a linear relation in log space, using a common slope while each ridge has its own zero point.
% \begin{equation}
% \mu_{ik}=a\,(x_i-x_{\rm p})+b_k, \qquad
% x_i=\log_{10}\!\left(\frac{P_i}{\rm d}\right),
% \end{equation}
% where $y_{ik}$ denotes either $\log_{10}(L_i/L_\odot)$ or $M_{G,0,i}$ for star $i$ assigned to ridge $k$, and $x_{\rm p}$ is the median $\log_{10}P$ of the combined ridge sample. Subtracting this numerical pivot reduces the covariance between the fitted slope and zero points without changing the resulting relations. 
% The two ridges share a common slope, $a$, while each has its own zero point, $b_k$, corresponding to the predicted value of $y$ at $x_{\rm p}$, and intrinsic scatter $s_k$. 
Each measurement uncertainty was added in quadrature to the corresponding intrinsic scatter. Quoted statistical intervals are the 16th--84th percentiles of 100 bootstrap resamples. 

The fitted $G$-band relations are
\begin{align}
\rev{M_{G}^{\rm F}} &=
\left(-3.041^{+0.012}_{-0.015}\right)
\log_{10}\!\left(\frac{P}{\rm d}\right)
-1.542^{+0.014}_{-0.017},\nonumber\\
\rev{M_{G}^{\rm S}} &=
\left(-3.041^{+0.012}_{-0.015}\right)
\log_{10}\!\left(\frac{P}{\rm d}\right)
-2.486^{+0.017}_{-0.019},
\label{eq:pl_fit}
\end{align}
with intrinsic scatters of $0.252$ and $0.320$\,mag, respectively. The shared slope is broadly consistent with the approximately $-3$ slopes obtained in previous radial-mode calibrations, although direct comparisons certainly depend on the passband and sample definition \citep{McNamara2011Deltaa, Ziaali2019Period, Jayasinghe2018ASASSN, Barac2022Revisiting, Guo2025Investigating, Jia2025Period,Murphy2026Population}.

Fitting luminosity in the same way gives
\begin{align}
\log_{10}\!\left(\frac{L_{\rm F}}{L_\odot}\right)
 &=\left(1.204^{+0.007}_{-0.004}\right)
 \log_{10}\!\left(\frac{P}{\rm d}\right)
 +2.476^{+0.008}_{-0.004},\nonumber\\
\log_{10}\!\left(\frac{L_{\rm S}}{L_\odot}\right)
 &=\left(1.204^{+0.007}_{-0.004}\right)
 \log_{10}\!\left(\frac{P}{\rm d}\right)
 +2.858^{+0.009}_{-0.006},
\label{eq:pl_luminosity_fit}
\end{align}
with intrinsic scatters of $0.104$ and $0.127$\,dex. We show this fit in Fig.~\ref{fig:pl}. Equivalently, adopting $M_{\rm bol,\odot}=4.74$, the bolometric relations have a shared slope of $-3.009$ and zero points of $-1.450$ and $-2.404$\,mag. The fitted zero-point separation corresponds to a horizontal displacement $\Delta\log P=-0.310$\,dex in \rev{$M_G$}, or $P_{\rm S}/P_{\rm F}=0.489^{+0.003}_{-0.001}$; the luminosity fit independently gives $0.482^{+0.002}_{-0.001}$. Again, these ratios are much closer to a third- or fourth--overtone than a first-overtone sequence.
%, but should not be interpreted as period ratios measured from two modes in the same star without manual identification of fundamental and overtone pulsations.  
% We further point out that the quoted uncertainties on our fit are certainly a lower bound on the true uncertainty.

\subsection{Rotational broadening and proximity to the radial-mode ridges}
\label{ssec:ridge_rotation}

\begin{figure*}[t!]
\centering
\includegraphics[width=0.49\linewidth]{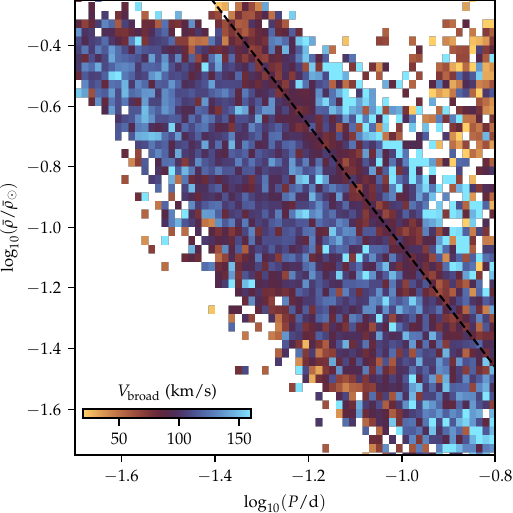}
\includegraphics[width=0.49\linewidth]{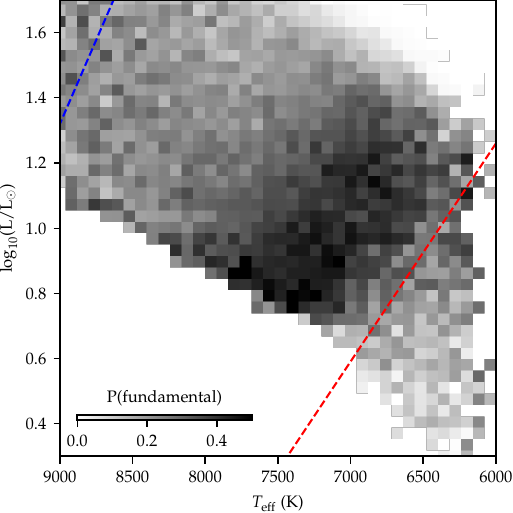}
\caption{\textit{Left:} The period-density relation colored by median \vbroad, with the fundamental ridge marked by a dashed black line. \textit{Right:} The \dsct\ sample in the H--R diagram, colored by probability of belonging to the fundamental radial ridge in the period--density relation. The dashed lines are the instability strip from Sec.~\ref{ssec:edges}.}
\label{f:dist_from_pl}
\end{figure*}

Color-coding the period--density relation by the \textit{Gaia} DR3 line-broadening parameter $v_{\rm broad}$ (Figure~\ref{f:dist_from_pl}, left) reveals that stars lying closest to the fundamental ridge has, on average, the lowest $v_{\rm broad}$, and that the distance of a star from the nearest ridge tends to increases with its rotational broadening. We interpret this correlation as the combined signature of two rotation-related effects that act, respectively, on the mode that is measured, on the density that is inferred, and on the pulsation frequency itself.

Firstly, the main ridge is populated by stars in which the fundamental radial mode is the dominant mode. The high-amplitude $\delta$~Scuti stars, which obey the period--density relation most tightly, are predominantly slow rotators \citep[$v\sin i \lesssim 30~{\rm km\,s^{-1}}$;][]{Breger2000Scuti}, whereas rapid rotators are overwhelmingly low-amplitude, multiperiodic pulsators dominated by non-radial modes \citep{Royer2007Rotational}. For the latter, the highest-amplitude peak often does not correspond to the fundamental radial mode, so the period shown in the diagram does not lie on that ridge. A substantial part of the trend with $v_{\rm broad}$ is therefore a mode-selection effect rather than a dynamical displacement.

Relatedly, the mean density is inferred from the luminosity and effective temperature under the assumption of a spherical, non-rotating star. Rapid rotation breaks this assumption, such that centrifugal flattening reduces the mean stellar density \citep{murphy2026quantifying} and gravity darkening renders the apparent $L$ and $T_{\rm eff}$ dependent on the unknown inclination \citep{vonZeipel1924Radiative, Lara2011Gravity}, with a star viewed pole-on appearing hotter and more luminous than the same star seen equator-on. Because the inclination is effectively random across the sample, rotation introduces an inclination-dependent scatter that can reach $\sim0.1$~dex in $\log\bar\rho$ at moderate rotation rates, which is comparable to the $\sim0.12$~dex separation between the fundamental and first-overtone ridges. This effect scatters rapid rotators to either side of a ridge rather than displacing them coherently \citep{garciahernandez2024back}.

Finally, rotation perturbs the pulsation frequencies directly. Centrifugal distortion inflates the star and lowers its true mean density, lengthening the radial periods, while the Coriolis force shifts and splits the non-radial modes \citep{Aerts2010Asteroseismology, Suarez2006Effects, Guo2024Oscillation}. These effects are second order in the rotation rate, $\propto(\Omega/\Omega_{\rm crit})^{2}$, and displace a star from the non-rotating constant-$Q$ relation by up to a few hundredths of a dex at the rotation rates typical of the sample.

These two effects combine to produce the trends with $v_{\rm broad}$ seen in the left panel of Figure~\ref{f:dist_from_pl}.

% These three mechanisms should make measurable predictions. The inferred-density bias increases the variance of the perpendicular distance while leaving its mean near zero, whereas the intrinsic frequency perturbations produce a coherent offset toward longer period and lower density. The distribution of perpendicular distance as a function of $v_{\rm broad}$ therefore discriminates between them. We find \todo{that something happens}. 
% We confirm that the trend persists at fixed $T_{\rm eff}$, since $v_{\rm broad}$ carries known temperature- and brightness-dependent systematics and is least reliable at low broadening \citep{Fremat2023Gaia}.

\subsection{Stars near the fundamental-mode ridge}
\label{ssec:fund_location}

We now use the probability of belonging to the fundamental ridge as an empirical diagnostic of where such stars lie within the instability strip. In Sec.~\ref{sec:pdensity}, we fitted a sum of Gaussians to determine the probability of each star belonging to either the fundamental, secondary, or background distributions. The right panel of Figure~\ref{f:dist_from_pl} shows the H--R diagram of the \dsct\ sample binned according to their probability of belonging to the fundamental ridge. The stars more likely to belong to the ridge are not distributed uniformly across the strip; they are concentrated {towards the cool, low-luminosity corner near the red edge}. The remaining stars appear to {populate the hotter and more luminous region towards the blue edge}. This pattern is consistent with the rotation picture developed above, because the cool, low-luminosity part of the strip contains many of the slowest rotators and HADS stars. It also implies that a PL relation calibrated from ridge-proximate stars samples a restricted region of the instability strip, a caveat for distance-scale applications. We stress that ridge proximity alone does not uniquely identify the radial fundamental: non-radial modes and rotationally shifted modes can overlap the ridge, while independently identified radial-fundamental pulsators may lie elsewhere in the diagram.

\section{\dsct\ pulsators in open clusters}
\label{sec:clusters}

In Sec.~\ref{ssec:freq_hrd} we examined how the mean frequency of \dsct\ stars decreases as the star evolves. Such a relation naturally invites us now to investigate how the mean pulsation frequencies of \dsct\ stars change within clusters.  Open clusters offer a powerful laboratory for the study of \dsct\ pulsations. Cluster members share a common age, distance, and initial chemical composition to within small dispersions, allowing the intrinsic variations in pulsation properties at fixed stellar parameters to be isolated from the population-level scatter that dominates the field-star sample analysed in Section~\ref{sec:results}. Several recent studies have exploited \kepler, K2 and \tess\ data to characterize the \dsct\ populations of individual clusters \citep[e.g.,][I. Berry et al., submitted]{Rodriguez2020, Murphy2022Five, Bedding2023TESS, Ortega2023Dating, Palakkatharappil2023, Li2024, Murphy2024Scuti, Berry2025Discovery, Dileep2025, Mwanzia2025Stellar, Qin2026Precise, Sores-Furtado2026, Zhou2025}, but a homogeneous analysis across the full set of clusters observable by \tess\ has so far not been completed. With \textit{Gaia} DR3 we now have cluster membership catalogues based on \textit{Gaia} astrometry which provide a homogeneous, all-sky foundation for population-level cluster asteroseismology.

\subsection{Cluster cross-matching}
\label{ssec:cluster_xmatch}

We cross-matched our full target sample against the \textit{Gaia} DR3 open cluster catalogue of \citet{Hunt2023Improving}, which provides homogeneous membership probabilities, ages, distances, and metallicities for 7167 clusters identified from \textit{Gaia} astrometry. We retained stars with membership probability $p_{\rm mem} > 0.7$ to minimize contamination from field stars, and we further restricted the cluster sample to those with at least five \dsct\ candidates within our analysis sample. This yielded \rev{69 clusters containing 579 \dsct\ pulsators spanning ages from $\log(t/{\rm yr}) = 6.84$ to $9.16$. The twenty clusters with the most \dsct\ members are listed in Table~\ref{tab:clusters}, and we make the full table and the member list available in machine-readable format.}

\begin{table}[t]
\centering
\begin{tabular}{lrrrr}
\toprule
Name & $N$ & $N_{\delta\,\mathrm{Sct}}$ & Age (Myr) & $\langle \nu_{\rm max} \rangle$ (\cd) \\
\midrule
NGC 3532 & 3414 & 40 & $238^{+218}_{-93}$ & $21.2 \pm 1.3$ \\
NGC 2360 & 1188 & 28 & $1451^{+703}_{-492}$ & $17.2 \pm 0.7$ \\
NGC 5822 & 804 & 27 & $1040^{+643}_{-330}$ & $16.9 \pm 0.9$ \\
NGC 2516 & 3784 & 25 & $125^{+127}_{-52}$ & $25.8 \pm 1.9$ \\
NGC 2423 & 571 & 19 & $1182^{+771}_{-348}$ & $14.8 \pm 0.9$ \\
NGC 6475 & 1427 & 16 & $171^{+163}_{-59}$ & $25.8 \pm 2.1$ \\
NGC 2099 & 2424 & 14 & $486^{+246}_{-165}$ & $12.2 \pm 0.6$ \\
NGC 1647 & 982 & 13 & $117^{+91}_{-42}$ & $34.2 \pm 2.6$ \\
NGC 6940 & 864 & 12 & $770^{+431}_{-242}$ & $18.9 \pm 1.6$ \\
NGC 2447 & 1043 & 11 & $740^{+449}_{-289}$ & $16.3 \pm 1.4$ \\
NGC 2632 & 1314 & 11 & $346^{+275}_{-145}$ & $21.9 \pm 1.7$ \\
NGC 6405 & 974 & 11 & $52^{+39}_{-22}$ & $27.2 \pm 2.9$ \\
ASCC 90 & 184 & 10 & $409^{+240}_{-148}$ & $20.5 \pm 2.0$ \\
NGC 2287 & 948 & 10 & $154^{+130}_{-58}$ & $24.5 \pm 3.1$ \\
NGC 6811 & 435 & 10 & $1078^{+698}_{-314}$ & $18.8 \pm 1.3$ \\
OC 0588 & 409 & 10 & $232^{+164}_{-95}$ & $19.8 \pm 2.4$ \\
Alessi 6 & 203 & 9 & $366^{+210}_{-127}$ & $14.7 \pm 2.4$ \\
NGC 2477 & 3543 & 9 & $701^{+358}_{-212}$ & $13.2 \pm 1.0$ \\
NGC 1662 & 384 & 8 & $345^{+243}_{-128}$ & $19.7 \pm 2.7$ \\
NGC 2354 & 399 & 8 & $1101^{+762}_{-328}$ & $12.6 \pm 0.9$ \\
\bottomrule
\end{tabular}
\caption{The 20 clusters with the most \dsct\ members. Here, $N$ is the number of cluster members in \citet{Hunt2023Improving}, $N_{\delta\,\rm Sct}$ is the number of members with $p_{\rm mem} > 0.7$ and a classifier score above 0.9, and $\langle \nu_{\rm max} \rangle$ is their mean pulsation frequency with its standard error (Section~\ref{ssec:freq_age}). Ages are from \citet{Hunt2023Improving}. The full table of 69 clusters, including the numbers of flagged blends and pre-main-sequence candidates, and the list of \dsct\ members are available in machine-readable form.}
\label{tab:clusters}
\end{table}

\subsection{Mean pulsation frequency as an age tracer}
\label{ssec:freq_age}

\begin{figure}
\centering
\includegraphics[width=\linewidth]{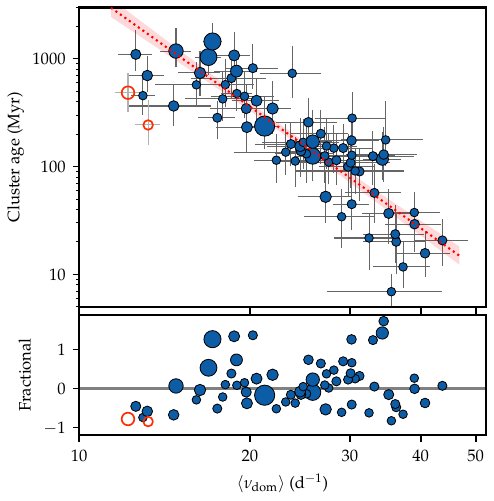}
\caption{Cluster age as a function of mean pulsation frequency. Each point represents a single cluster, with the marker size scaled by the number of \dsct\ members and the error bars showing the standard error on the mean frequency and the catalogue age uncertainty. The dotted line shows the best-fit linear relation (excluding the two clusters shown as red symbols), and the lower panel shows the fractional residual.}
\label{f:cluster_age}
\end{figure}

% For each cluster we compute the mean dominant pulsation frequency $\langle \nu_{\rm dom} \rangle$ of its \dsct\ members, weighted by the amplitude of the mode. Figure~\ref{f:cluster_age} shows $\langle \nu_{\rm dom} \rangle$ against cluster age. There is an obvious correlation where clusters younger than $\sim 100$\,Myr have mean pulsation frequencies of $\sim 30$--$50$\,\cd, while clusters older than $\sim 1$\,Gyr cluster at $\sim 10$--$20$\,\cd. 

For each cluster we computed the mean pulsation frequency $\langle \nu_{\rm max} \rangle$ of its \dsct\ members. For an individual star, we defined this to be the amplitude-weighted mean of its prewhitened frequencies \citep{BarceloForteza2020Unveiling, Murphy2024$d$}, and the cluster value is the unweighted mean over its members. Figure~\ref{f:cluster_age} shows cluster age against $\langle \nu_{\rm max} \rangle$. There is an obvious correlation where clusters younger than $\sim 100$\,Myr have mean pulsation frequencies of $\sim 30$--$50$\,\cd, while clusters older than $\sim 1$\,Gyr cluster at $\sim 10$--$20$\,\cd. We regressed $\log t$ on $\log \nu_{\rm max}$, which is the direction in which the relation could be used (see below), and obtained
\begin{equation}
\label{eq:freq_age}
\log\!\left(\frac{t}{\rm yr}\right) = -3.75 \small( \pm 0.36)\,\log\!\left(\frac{\langle \nu_{\rm max} \rangle}{{\rm \cd}}\right) + 13.43(\pm 0.51),
\end{equation}
with an intrinsic scatter of $\sigma_{\log t} \approx 0.09$\,dex. The fit carries the uncertainty on both coordinates together with an intrinsic-scatter term. Two clusters (red circles in Figure~\ref{f:cluster_age}), Ruprecht~98 and NGC~2099, were rejected by a $3\sigma$ clipping, leaving 67 clusters in the fit. The physical origin of this correlation is well understood from the fact that mean stellar density decreases with age, such that the average frequency also decreases.

% The physical origin of this correlation is well understood from stellar evolution. As a \dsct\ star evolves on the main sequence, its mean density decreases and the frequencies of its acoustic modes, which scale as $\nu \propto \bar{\rho}^{1/2}$ (Eq.~\ref{eq:pulsation_constant}), fall correspondingly. In a coeval cluster, the most massive stars currently within the instability strip have arrived there from hotter spectral types and are therefore close to the terminal-age main sequence. The mass of these instability-strip stars is set by the cluster age, with older clusters hosting progressively lower-mass \dsct\ pulsators that have correspondingly longer main-sequence lifetimes and lower densities. The mean pulsation frequency of the \dsct\ population is therefore a tracer of the mass at which the instability strip is currently being populated, and hence of the cluster age.

% We also briefly inspect the mean spacing between the modes of a handful of targets, which also appear to show a strong correlation with the cluster age. However, manual measurement of the large-spacing for all of these targets is beyond the scope of this paper and we leave it to future work.

Much of the substantial scatter about this relation presumably arises from the star-to-star variation of which modes are excited within each cluster, as seen with \tess\ in the Pleiades \citep{Bedding2023TESS}, NGC~2516 \citep{Li2024}, NGC~3532 \citep{Berry2025Discovery}, and the Cep-Her complex \citep{Murphy2024$d$}.
Nevertheless, the empirical relation in Eq.~\ref{eq:freq_age} offers a practical age indicator for young to intermediate-age clusters that complements traditional isochrone fitting. Isochrone-based ages rely on accurate photometric calibration, well-constrained extinction, and reliable main-sequence-turnoff identification, all of which can be problematic for sparsely populated or heavily reddened clusters. The \dsct\ mean-frequency age, by contrast, requires only that several \dsct\ pulsators be identified within the cluster and that their dominant frequencies be measured, both of which are straightforward with \tess\ 10-min FFI data. For an arbitrary cluster, the relation returns an age to $\sigma_{\log t} \approx 0.2$\,dex, which is a factor of $1.5$, so that a $500$\,Myr cluster is placed between $330$ and $760$\,Myr. This is comparable to the median uncertainty of $0.23$\,dex on the catalogue ages themselves, and to the precision achieved by careful isochrone fitting for similar clusters \citep{Bossini2019Age, Cantat-Gaudin2020Painting}.

% The intrinsic scatter of {$\sim 0.057$\,dex} corresponds to an age uncertainty of {$\sim X\%$}, comparable to that achieved by careful isochrone fitting for similar clusters \citep{Bossini2019Age, Cantat-Gaudin2020Painting}.

% We note that the relation is calibrated against the isochrone ages it is intended to complement, so it inherits any systematic error in those ages; it is best understood as a fast age estimator rather than an independent age scale, until detailed stellar modeling can be performed. 

This relation is defined by the mean dominant frequency, $\langle \nu_{\rm max} \rangle$, and is therefore sensitive to small-number statistics in clusters with few \dsct\ members. We have restricted our analysis to clusters with at least five identified pulsators to mitigate this but the youngest and oldest clusters in our sample, where \dsct\ stars are rare, carry correspondingly larger uncertainties. Secondly, metallicity is expected to influence both the location of the instability strip and the pulsation frequencies of stars within it {\citep{Murphy2019Gaiaderived, Murphy2026Population}}. Thirdly, rapidly rotating \dsct\ stars exhibit modified pulsation frequencies due to rotational distortion (Section~\ref{ssec:ridge_rotation}); the impact of rotation on cluster-mean frequencies will require further investigation with a larger sample, which we identify as an avenue for future work.

% \subsection{\texorpdfstring{\rev{Pre-main-sequence cluster candidates}}{Pre-main-sequence candidates}}
\subsection{Pre-main sequence cluster candidates}
\label{ssec:pms}

\rev{Stars of 1.5--2.5\,\Msun\ cross the instability strip during their pre-main-sequence (PMS) contraction towards the ZAMS, and a number of PMS \dsct\ pulsators are known \citep{Zwintz2008Comparing, Zwintz2014Echography, Murphy2021Precise, Steindl2021Tidally}. Their pulsation frequencies increase as they contract making them a direct and useful probe of the earliest evolutionary stages \citep{Zwintz2014Echography, Zwintz2022PreMain, gautam2026grid}. PMS and main-sequence \dsct\ stars overlap in the H--R diagram, however, and cannot be separated using \teff\ and $L$ alone \citep{Zwintz2008Comparing}.}

\rev{However, membership of a young cluster provides a useful age constraint, allowing us to identify potential PMS \dsct\ targets. For each \dsct\ cluster member, we compared the age of its host cluster with the time needed for a star of its \textsc{isoclassify} mass to reach the ZAMS. We took these arrival times from the solar-metallicity, non-rotating MIST tracks \citep{Choi2016MESA, Dotter2016MESA}, in which a star reaches the ZAMS after roughly 14\,Myr at 1.5\,\Msun, 7.5\,Myr at 2.0\,\Msun, and 4.5\,Myr at 2.5\,\Msun. The \textsc{isoclassify} masses are derived from main-sequence isochrones, so they are biased for genuine PMS stars. A contracting A star lies above the ZAMS at close to its final luminosity, and main-sequence isochrones interpret this position as an evolved star of lower mass. The resulting underestimate of the mass, by up to ${\sim}20\%$, lengthens the expected arrival time. This bias therefore cannot cause a genuine PMS star to be missed, and it does not affect the masses of main-sequence members. Of the 1515 \dsct\ stars with $p_{\rm mem} > 0.7$, 74 stars belong to clusters whose median age is younger than the arrival time.  They include members of the young clusters NGC~2244, NGC~6383, and Mamajek~1. Their dominant frequencies (median 22\,\cd) are lower than those of the remaining \dsct\ members of clusters younger than 100\,Myr (median 32\,\cd), even though the two groups have the same median \teff. This is the sense expected for stars that have not yet contracted onto the ZAMS \citep{Zwintz2014Echography}. We flag all 74 candidates in Table~\ref{tab:params} and in the machine-readable list of cluster members.}

\section{Conclusions}
\label{sec:conclusions}

We have presented a large, uniform asteroseismic survey of A- and F-type stars, comprising \samplesize\ stars in and around the classical \dsct\ instability strip observed by \tess\ at 10-minute full-frame image cadence over most of the sky. The $72$\,\cd\ Nyquist frequency greatly reduces the aliasing that affects 30-minute data, while \textit{Gaia} DR3 parallaxes and \textsc{isoclassify} fits provide homogeneous stellar parameters across the sample. Comparisons with interferometric temperatures and angular diameters support the adopted temperature and radius scales, with the residual scatter described in Section~\ref{ssec:interf_validation}. Our results are as follows.

\begin{itemize}
\item The \dsct\ instability strip is not pure at the detection threshold of this survey, but the apparent impurity is strongly magnitude-dependent. The peak pulsator fraction is {$\sim 30\%$} for the full sample and rises to {$\sim 80\%$} for $T_{\rm mag} < 9$, where our amplitude sensitivity is greatest. Detection sensitivity therefore accounts for much of the difference between our full-sample fraction and the {$\sim 60\%$} inferred from \kepler\ \citep{balonaKeplerObservationsScuti2011, Murphy2019Gaiaderived}. Even at the bright end, the detected fraction does not reach unity, confirming that a genuine population of non-\dsct\ stars pervades the strip.

\item \dsct\ pulsators are systematically faster rotators than non-pulsating stars at the same position in the H--R diagram, with the pulsator fraction increasing monotonically with the \textit{Gaia} \vbroad\ parameter. This confirms the findings of \citet{Murphy2024Scuti, Gootkin2024New, Wang2025Rotation} with a larger sample and unambiguous frequency coverage. A population of \dsct\ pulsators with $\vbroad < 50$\,km\,s$^{-1}$ persists, larger than can be explained by inclination effects alone, and requires spectroscopic follow-up.

\item We refine the \dsct\ period--density and period--luminosity relations, finding results largely consistent with previous determinations \citep{McNamara2011Deltaa, Ziaali2019Period, Barac2021Revisiting}. Recast as a period--density relation, two approximately parallel ridges are resolved with a ratio of pulsation constants of 0.578, corresponding to a likely combination of excited overtone modes.

\item Stars lying closest to the radial-mode ridge have systematically lower \vbroad, and the distance from the ridge increases with rotational broadening. We attribute this to a combination of mode selection (rapid rotators are dominated by non-radial modes), inclination-dependent bias in the inferred mean density from gravity darkening, and direct second-order rotational perturbation of the frequencies. These fundamental-mode pulsators occupy a {cooler, less luminous} subregion of the instability strip.

\item The dominant pulsation frequency varies systematically across the instability strip, decreasing from {$\sim 50$--$70$\,\cd} near the blue edge to {$\sim 10$--$20$\,\cd} near the red edge, as expected from the scaling of acoustic frequencies with mean stellar density. 
% We find that \todo{$X\%$} of \dsct\ pulsators have their dominant mode above the \kepler\ long-cadence Nyquist limit, quantifying the population for which previous long-cadence surveys provided only aliased measurements.

\item The mean pulsation frequency, $\langle \nu_{\rm max} \rangle$, of \dsct\ stars in open clusters correlates with cluster age, with $\log\langle\nu_{\rm dom}\rangle \propto$ {$a \log t$}. This provides a new ensemble-asteroseismic age indicator that complements traditional isochrone-based methods.
\end{itemize}

Several avenues for future work follow directly. Spectroscopic follow-up of the slowly-rotating \dsct\ pulsators will clarify whether these are physically distinct from the chemically peculiar Am stars that dominate the slow-rotator population in the same temperature range. The asymmetric mode splitting predicted by \citet{Guo2024Oscillation} provides a path towards mode identification in rapidly rotating \dsct\ stars that should now be testable on the larger sample assembled here. Finally, the cluster mean-frequency--age relation invites extension to the older Gyr-scale clusters where \dsct\ pulsators are rare, as well as replacement of $\langle \nu_{\rm max} \rangle$ with a per-star $\Delta\nu$, and a multi-cluster ensemble analysis that may simultaneously constrain age, metallicity, and helium abundance. The continued operation of \tess\ in faster cadence, together with the future delivery of \textit{Gaia} DR4 astrometry and PLATO photometry \citep{Rauer2025PLATO}, will further improve our understanding of \dsct\ variability.

\begin{acknowledgments}

{DRH acknowledges support from TESS GI programs {80NSSC24K0506}, {80NSSC25K7910}, and NASA ADAP {80NSSC24K0621}.}
{SJM was supported by the Australian Research Council through Future Fellowship FT210100485.}
{TRB and PM were supported by the Australian Research Council through Laureate Fellowship FL220100117.}
{D.H. and I.B. acknowledge support from the National Aeronautics and Space Administration (80NSSC25K7155).} 
{MK acknowledges The Kavli Foundation for their financial support in the framework of the Kavli Scholarship given to MK from 25/9/2023–30/9/2026.}
{HK acknowledges support from National Science Foundation (NSF) Research Experience for Undergraduates (REU) Grant No. 2548017.}
{This paper includes data collected with the TESS mission, obtained from the MAST data archive at the Space Telescope Science Institute (STScI). Funding for US Institutions for the TESS mission is provided by the NASA Explorer Program. STScI is operated by the Association of Universities for Research in Astronomy, Inc., under NASA contract NAS 5–26555. This work has made use of data from the European Space Agency (ESA) mission \textit{Gaia} (\url{https://www.cosmos.esa.int/gaia}), processed by the \textit{Gaia} Data Processing and Analysis Consortium (DPAC).}

The playlist made while writing this paper can be found \href{https://open.spotify.com/playlist/57M1XRIp3eta5N0TS2bP8S?si=b18db10ab74445e0}{here}. 

\end{acknowledgments}

\facilities{TESS, Gaia}

\software{astropy \rev{\citep{astropycollaborationAstropyCommunityPython2013, price-whelanAstropyProjectBuilding2018, astropy2022}},
          numpy \citep{Harris2020Array},
          scipy \citep{2020SciPy-NMeth},
          matplotlib \citep{Hunter2007Matplotlib},
          lightkurve \citep{GeertBarentsen2019KeplerGO},
          isoclassify \citep{Huber2017Asteroseismology},
          mwdust \citep{Bovy2016Galactic},
          TGLC \citep{Han2023TESSGaia}
          }

\appendix

\section{De-reddening Gaia photometry}
\label{app:extinction}

The \texttt{isoclassify} posteriors provide line of sight extinction in the $V$ band, whereas our formulation of absolute magnitude requires extinction in the \textit{Gaia} $G$ band. The \textit{Gaia} passbands are broad enough that the ratio $k_X\equiv A_X/A_V$ is not a constant but rather depends on both the total extinction and on the intrinsic spectral energy distribution of the star, which, for our purposes is parametrised by the intrinsic color $(G_{\rm BP}-G_{\rm RP})_0$. 

Over the color and extinction range spanned by our sample, $k_G$ varies between $0.84$ and $0.97$, so we can not adopt a single fixed coefficient, which would bias $A_G$ by up to $0.06$\,mag. We therefore adopt the coefficients tabulated by \citet{Luhman2020} in their Table~8, who fit $k_{\rm G, BP, RP}$ as polynomials in $A_V$ and $(G_{\rm BP}-G_{\rm RP})_0$. of the form
\begin{equation}
    k_X = c_0 + c_1  x + c_2 A_V + c_3 xA_V + c_4 x^2 A_V + 5 A_V^2
      + c_6 x^3 + c_7 A_V^3 + c_8 x^2 A_V^2 ,
\qquad x \equiv (G_{\rm BP}-G_{\rm RP})_0 ,
\label{eq:kpoly}
\end{equation}
with separate coefficient $c_0 \ldots c_8$ for each band. These fits are calibrated against the optical--infrared extinction curve of \citet{Schlafly2016OPTICAL} evaluated at $x=0$ which corresponds to $R_V\simeq3.3$ rather than the usual $R_V=3.1$.

Note that Eq.~\ref{eq:kpoly} is a function of the intrinsic color, which is also not directly observed. Instead, we solve for it through an iteration regime. Starting from the observed color, 
\begin{equation}
x^{(n+1)} = (G_{\rm BP}-G_{\rm RP})
- \left[k_{\rm BP}\!\left(x^{(n)},A_V\right)
      - k_{\rm RP}\!\left(x^{(n)},A_V\right)\right] A_V ,
\qquad x^{(0)} = (G_{\rm BP}-G_{\rm RP}) ,
\label{eq:colour_iteration}
\end{equation}
and re-evaluate the BP and RP coefficients at the current estimate of the intrinsic color at each step. We find that the iteration converges to better than $10^{-4}$\,mag within six steps, however, for a large enough extinction, the fixed point can lie outside the range over which Eq.~\ref{eq:kpoly} is fitted and then the iteration diverges. We therefore restrict $x$ to the fitted range at each step and require $-0.2 < G_{\rm BP} - G_{\rm RP} < 1.5$ in the final sample selection. This second requirement removes stars for which the \texttt{isoclassify} $A_V$ posterior is too large to be consistent with an observed \textit{Gaia} color.

Having converged on $x$, the extinction in the $G$ band follows as $A_G=k_G(x,A_V)\,A_V$, and we take the statistical uncertainty on $A_G$ to be $k_G\sigma_{A_V}$, with $\sigma_{A_V}$ the width of the \texttt{isoclassify} $A_V$ posterior.

\section{\texorpdfstring{\rev{Crowding, blending, and Galactic latitude}}{Crowding, blending, and Galactic latitude}}
\label{app:blending}

\begin{figure*}[t]
\centering
\includegraphics[width=\linewidth]{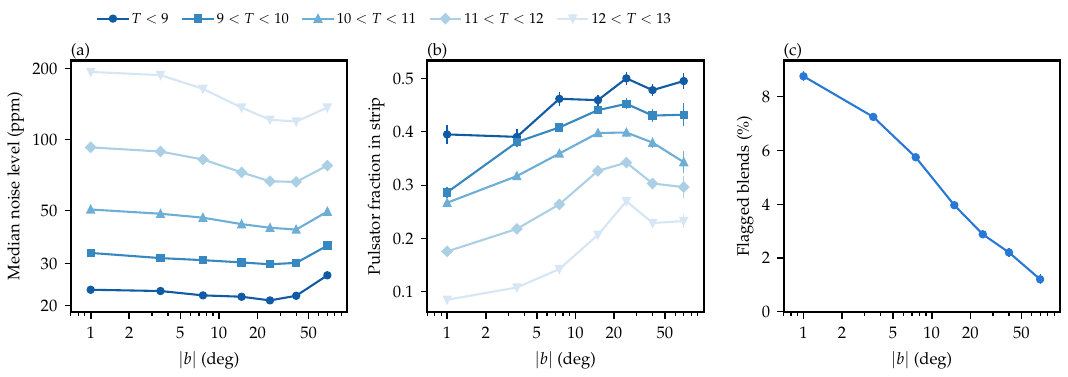}
\caption{\rev{Effect of crowding as a function of Galactic latitude. (a) Median noise level (95th percentile of the amplitude spectrum) of stars not classified as pulsators, in bins of \tess\ magnitude. (b) Fraction of stars within the instability strip of Section~\ref{ssec:edges} that are classified as \dsct\ pulsators, excluding flagged blends, in the same magnitude bins. (c) Percentage of classified \dsct\ stars flagged as probable blends.}}
\label{f:latitude}
\end{figure*}

\rev{Because 59\% of our parent sample lies within $10\arcdeg$ of the Galactic plane, we tested how crowding affects our photometry and results. The median TIC contamination ratio \citep{Stassun2018TESS} increases by approximately two orders of magnitude from the Galactic poles to the plane, such that the noise level at fixed magnitude rises towards the plane (Figure~\ref{f:latitude}a). The fraction of instability-strip stars that we detect as pulsators falls accordingly (Figure~\ref{f:latitude}b), which makes crowding a secondary source of incompleteness (alongside apparent magnitude).}

\rev{Because \textsc{TGLC} models the neighbouring stars as constant sources, their variability can contaminate into a target \citep{Han2023TESSGaia}. Of the targets within 1 pixel ($21\arcsec$) of a classified \dsct\ star, 92\% share its dominant frequency to within 0.005\,\cd. We therefore grouped all targets within 5 pixels of each other whose dominant frequencies agree to this tolerance. We took the member with the largest flux amplitude (the product of the relative amplitude and the \textit{Gaia} $G$-band flux) as the source of the signal and flagged the others as probable blends. This flags 8787 of the \ndsct\ classified \dsct\ stars (5.6\%), with the fraction rising from 1.2\% at $|b| > 50\arcdeg$ to 8.8\% at $|b| < 2\arcdeg$ (Figure~\ref{f:latitude}c). We point out that variable stars outside our target list can contaminate our targets in the same way, so this is a lower limit on the true blend fraction. We find that removing the flagged blends does not significantly change our results: the pulsator fraction within the strip changes from 20.9\% to 20.5\%, the median \vbroad\ of the pulsators changes by less than 1\,km\,s$^{-1}$.}

\begin{singlespace}
    \bibliographystyle{aasjournalv7}
    \bibliography{library}
\end{singlespace}

\end{document}